# Discovery of Itinerant Magnetic Domain Wall and Quasiparticle Boundary State in Spin-Density-Waves

Yining Hu[1†], Xu Wang[1†], Chen Chen[1,7]*, Qingle Zhang[1], Dongming Zhao[1], Tianzhen Zhang[1,8], Chenxi Wang[1], Qiang-Hua Wang[3,4], Donglai Feng[2]*, Tong Zhang[1,4,5,6]*

[1]Department of Physics, State Key Laboratory of Surface Physics and Advanced Material Laboratory, Fudan University; Shanghai 200438, China
[2]New Cornerstone Laboratory, Hefei National Laboratory, Hefei 230027, China
[3]National Laboratory of Solid State Microstructures & School of Physics, Nanjing University, Nanjing, 210093, China
[4]Collaborative Innovation Center for Advanced Microstructures; Nanjing 210093, China
[5]Hefei National Laboratory, Hefei 230088, China
[6]Shanghai Research Center for Quantum Sciences; Shanghai 201315, China
[7]Zhejiang Institute of Photoelectronics & Zhejiang Institute for Advanced Light Source, Zhejiang Normal University; Jinhua 321004, China.
[8]Interdisciplinary Materials Research Center, School of Materials Science and Engineering, Tongji University; Shanghai 201804, China.

†These authors contributed equally to this work
*Corresponding authors. cchen_physics@zjnu.edu.cn, dlfeng@ustc.edu.cn, tzhang18@fudan.edu.cn

**Conventional magnetic domain walls are characterized by reorientation of local spins. However, what occurs at the boundary of itinerant magnets is largely unknown. Here using spin-sensitive scanning tunneling microscopy, we investigated the microscopic domain wall structure of the spin-density-wave (SDW) state in a prototypical itinerant antiferromagnet—chromium (Cr). At the boundary of two incommensurate SDW domains, we found the spins undergo finite-scale decay rather than reorientation. This generates a double-Q SDW state, which is further evidenced by an accompanying second-order charge modulation. In the commensurate SDW domains, a clear SDW energy gap is observed. Interestingly, the screw dislocations induced "half" vortex and anti-vortex of SDW, paired by antiphase domain wall. The spin density vanished at such antiphase domain walls. Remarkably, for the first time we observed the SDW quasiparticle states at the boundary, resembling the Andreev bound states in superconductors. These unique SDW boundary structures can be viewed as consequences of local interference of two SDWs, either with different *Q* or reversed phases. Our findings thus reveal a new type of domain wall distinct to that of local moment magnetism, with a mechanism rooted in the itinerant nature of SDW.**

## I. INTRODUCTION

In materials exhibiting quantum orders, the presence of boundary or domain walls can generate new states and structures that are distinct from the interior. For instance, orders driven by interactions in an itinerant electron system often lead to energy gap opening near $E_F$ [1-3]. The boundaries or defects can disrupt the uniformity of order parameter and give rise to novel in-gap quasiparticle states or structures. Well-known examples are Andreev bound states (including Majorana quasiparticles) [4-8], topological defects such as vortices [9], and Josephson junctions [10], most of which are observed in superconductors so far.

Magnetic orders also exhibit rich structures at their domain boundaries, which are predominantly ascribed to the reorientation of local spins [11,12]. The Néel/Bloch type domain walls have been extensively studied for their importance in magnetization and spintronics applications (such as domain wall memory) [13,14]. On the other hand, magnetic orders can also arise from itinerant electrons [15], which poses a fundamental question: what occurs when an (pure) itinerant magnetism encountered its domain boundary? One might anticipate new spin textures, given the spins are not necessarily localized on atomic sites and their amplitude may not keep constant, thus the mechanism of domain wall formation could be different. Moreover, akin to other itinerant orders like superconductivity, quasiparticles or topological defect may be generated where the itinerant order parameter varies [16]. However, in contrast to the cases of local-moment magnetism, the boundary spin structure of itinerant magnetism remains largely unexplored so far, despite magnetization dependent electronic state was reported in some ferromagnetic domain walls [17-19].

In this work, we report a microscopic study on the boundary structure of spin density wave (SDW) state in Cr(001), an exemplary itinerant antiferromagnet. An SDW is usually driven by electron correlation and Fermi surface nesting [2,20], manifesting as real-space spin modulations. It can be formally viewed as electron-hole pairing and described by the mean-field BCS theory with a vector order parameter of $\eta = \boldsymbol{m}\, e^{i(\boldsymbol{Q}\cdot\boldsymbol{r}+\varphi_0)}$ ($m$ is the spin amplitude and $\boldsymbol{Q}$ is the wavevector) [1,21]. The bulk Cr with body-centered cubic lattice [Fig. 1(a)] is well-known to have an incommensurate SDW (*IC*-SDW) ground state [22,23], as characterized by neutron scattering [22-25], x-ray diffraction [26,27] and photoemission spectroscopy [28,29]. The projected Fermi surface of Cr [Fig. 1(b)] comprised a square-shaped electron pocket at Brillouin zone center (**Γ**) and a slightly larger hole pocket at the corners (**H** points**,** or **X/Y** points for different corners). This geometry yields two slightly different nesting vectors of $\boldsymbol{Q}_{\pm} = 2\pi(1\pm\delta)/a$ oriented along one of the <001> directions, which stabilize the long-period *IC*-SDW with $\boldsymbol{Q}_{IC\text{-SDW}} = (\boldsymbol{Q}_{+} - \boldsymbol{Q}_{-})/2$. The *IC*-SDW is accompanied by a charge density wave (CDW) with $\boldsymbol{Q}_{\mathrm{CDW}} = 2\boldsymbol{Q}_{IC\text{-SDW}}$, which connects the two folded bands at **Γ** [30,31]. At low temperatures (< 10K) the wavelength of *IC*-SDW is ≈6.0 nm with the spins oriented along $\boldsymbol{Q}$ [19], as illustrated in Fig. 1(c).

Although a cubic lattice has three equivalent <001> directions, Cr predominantly exhibits a single-Q *IC*-SDW state which breaks the lattice rotation symmetry. The multi-Q SDW (a superposition of different $\boldsymbol{Q}$) has been predicted to be energetically unfavorable [32,33]. Therefore, different single-$Q$ domains with orthogonal $\boldsymbol{Q}$ vectors will coexist [34,35]. Using spin-polarized scanning tunneling microscopy (SP-STM), we investigated the boundaries formed by different domains and SDW phase inversion. We observed various features that are distinct from the local-moment magnetism. Firstly, at the *IC*-SDW domain wall, the spin amplitude decays exponentially within a finite scale (SDW coherence length). A new "double-$\boldsymbol{Q}$" SDW is generated at the boundary, which induces a second-order charge modulation and determines the domain wall energy/width. Secondly, when $\boldsymbol{Q}$ is along out-of-plane, the SDW becomes commensurate (*C*) to lattice, and a clear SDW gap opens in tunneling spectrum, corresponding to commensurate nesting condition. In *C*-SDW domain, screw dislocations induced "half" vortex of SDW, leading to $\pm\pi$ phase change and generating antiphase domain walls. At the antiphase walls, we found both SDW gap and the spin density vanished, while the SDW in-gap states emerge. Our model calculation shows that this state resembles the Andreev bound state of superconductors with gap sign change. These unique boundary structures and SDW quasiparticle states highlight the itinerant nature of SDW, and represent a new type of magnetic domain walls in itinerant magnets.

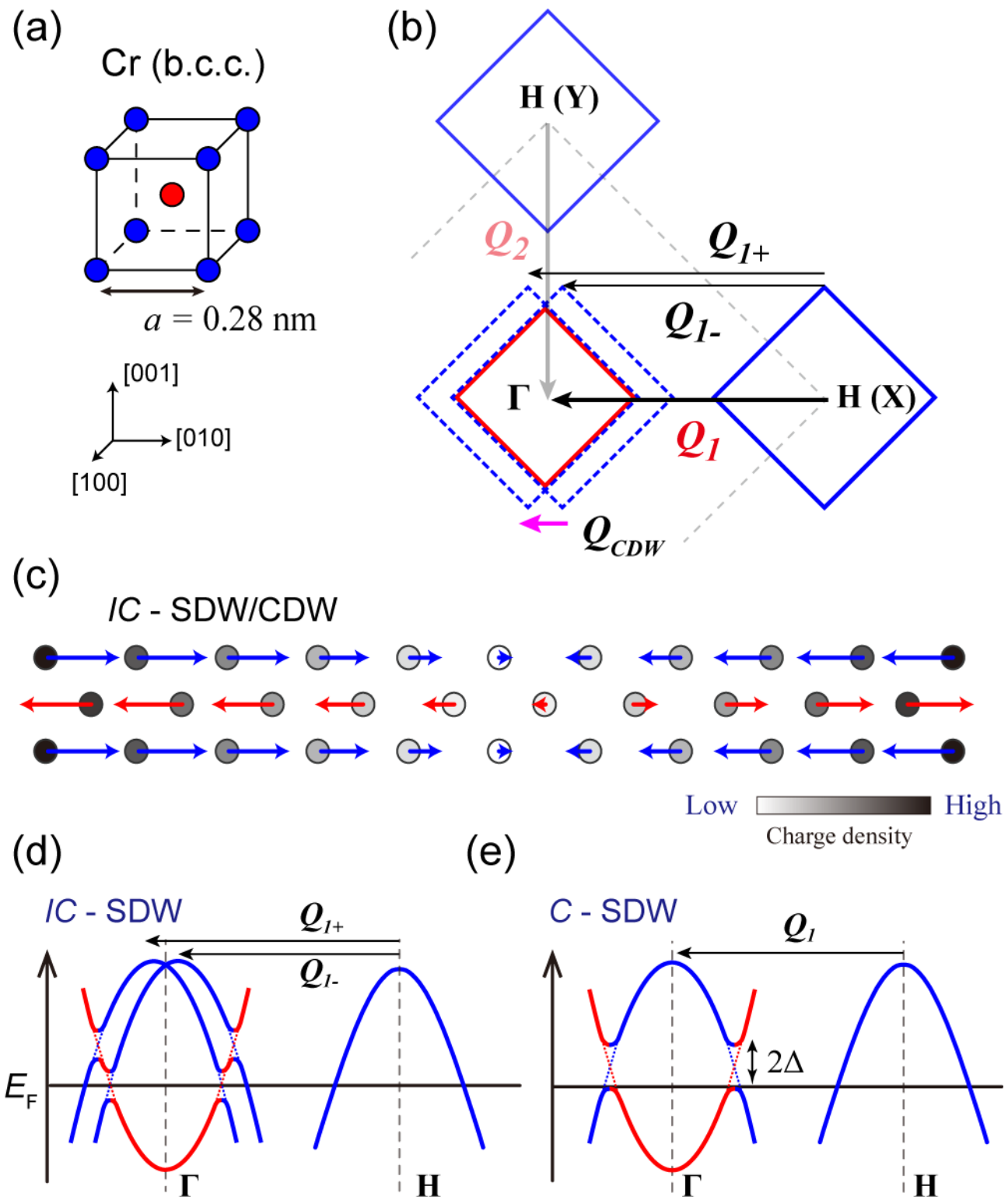

**FIG. 1. The structure, Fermi surface and SDW state of Cr.** (a) The lattice structure and (b) the (001) projected Brillouin zone of Cr. The electron and hole Fermi pockets (cross sections) are represented by red and blue squares, respectively. The nesting vectors $\boldsymbol{Q}_1$ and $\boldsymbol{Q}_{1\pm}$ are indicated by black arrows. $\boldsymbol{Q}_{CDW} = \boldsymbol{Q}_+ - \boldsymbol{Q}_-$ is the CDW wavevector. (c) The spin/charge structure of *IC*-SDW/CDW (below $T_{SF}$ =123K). (d, e) Sketches of band folding and energy gap opening for *IC*- and *C*- SDW, respectively.

## II. EXPERIMENTAL RESULTS

### A. Spin-polarized STM characterization of the *IC*- and *C*- SDW domains

The experiment was conducted in a cryogenic STM system at T = 4.5 K [see Method section]. Fig. 2(a) shows a typical topographic image of cleaned Cr (001) surface, featuring flat terraces with single-atomic layer height (inset). Using Fe-coated tip magnetized by a vector magnetic field, we can measure dI/dV maps with spin sensitivity along X, Y and Z directions (refer to [100], [010] and [001], respectively). Subsequently, pure spin contrast maps of each direction ($S_X$, $S_Y$, $S_Z$) were obtained by the relative difference of maps taken under $B_{\pm X}$, $B_{\pm Y}$ and $B_{\pm Z}$, respectively (see supplementary material part-**I** for more details).

A prior study has reported in-plane *IC*-SDW domains with $\boldsymbol{Q}$ along X or Y, as projections from the intrinsic bulk IC-SDW [36]. Here we find that the $\boldsymbol{Q}$ along Z (out-of-plane) domain also exists in the same crystal. Fig. 2(b) is the $S_Z$ map taken on the same area shown in Fig. 2(a), which displays two domains with different spin contrast. In domain on the right, adjacent terraces have opposite $S_Z$ component (the inversion of $S_Z$ in single terrace is discussed later); while no $S_Z$ contrast is observed in the left domain. Meanwhile, the $S_X$ map [Fig. 2(c)] clearly reveals in-plane *IC*-SDW in the left domain but no contrast on the right. This indicates the right domain has a $\boldsymbol{Q}$ along Z. We further found the $S_Z$ component of right domain does not show long-period modulation over a series of terraces (see supplementary material part **II**). This evidences the SDW became commensurate to lattice when $\boldsymbol{Q}$ is along Z. Hereafter, we refer to in-plane *IC*-SDW as $\boldsymbol{Q}_{1\pm}/\boldsymbol{Q}_{2\pm}$ domain, and the out-of-plane SDW as $\boldsymbol{Q}_3$ domain.

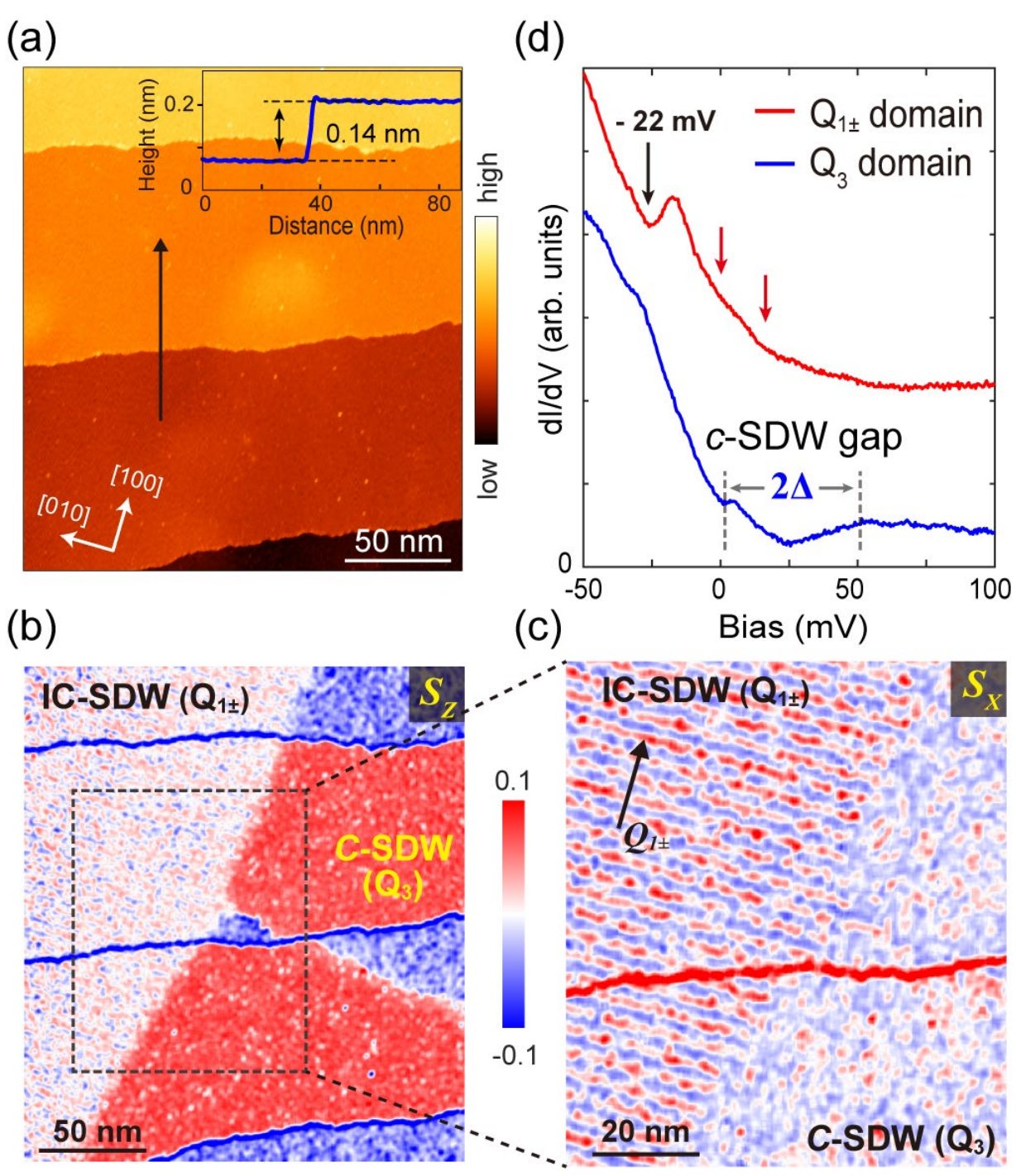

**FIG. 2. SP-STM characterization of SDW domains in Cr with different $\boldsymbol{Q}$ orientations.** (a) Topographic image of Cr (001) surface ($V_b$ = 1V, $I$ = 30 pA). (b) The $S_Z$ map of the area shown in (a). (c) $S_X$ map taken in the dashed box in (b). (d) Typical dI/dV spectra taken on $\boldsymbol{Q}_{1\pm}$ and $\boldsymbol{Q}_3$ domain (setpoint: $V_b$ = -50mV, $I$ = 60 pA). A DOS dip at $V_b \approx$ -22 mV is related to CDW [36].

Then we carried out tunneling conductance (dI/dV) measurement on different SDW domains. Fig. 2(d) shows the dI/dV spectra near Fermi level taken on $\boldsymbol{Q}_{1\pm}$ and $\boldsymbol{Q}_3$ domains (See Fig. S3 for large energy scale spectra). As an itinerant order, SDW is expected to open single-particle energy gap(s) near $E_F$, however the gap could have different forms for *IC* and *C* cases. Fig. 1(d) illustrates the *IC*-SDW induced band folding near $\boldsymbol{\Gamma}$ point. The two folded hole bands (blue) by $\boldsymbol{Q}_{\pm}$ only open partial gaps at their intersections with the electron band (red). The spectrum of $\boldsymbol{Q}_{1\pm}$ domain shows some weak kinks near $E_F$ (indicated by red arrows), possibly originating from these partial gaps (the dip at $V_b$ = -22 mV is associated with incommensurate CDW [36]). In contrast, the spectrum of $\boldsymbol{Q}_3$ domain is rather different. It displays a gap-like feature centered at $V_b \approx 25$ mV with a full width of ≈50 mV, and the dip at -22 mV disappeared. This gap can be explained by the band folding induced by *C*-SDW. As depicted in Fig. 1(e), since the hole pocket is slightly larger than the electron pocket, a direct folding from **H** to **Γ** by $\boldsymbol{Q}_1$ will open a full gap above $E_F$. Hence, both the spin textures and tunneling spectrum suggest an *IC* to *C* transition occurs when $\boldsymbol{Q}$ switched from in-plane to out-of-plane. This is consistent with previous neutron scattering studies which evidenced commensurate SDW domains coexisting with the *IC*-SDW [24,25], and ARPES studies on Cr (110) have reported a *C*-SDW gap opening above $E_F$ [28,29].

We note various early STM studies also observed inter-layer spin reversion (i.e., a layer-wise AFM state) on Cr(001) [37-43], but with a spin polarization lies in-plane. This contrasts with the out-of-plane spin polarization in $\boldsymbol{Q}_3$ domain here. Moreover, the SDW gap and *IC*-SDW domain have not been observed in previous studies. These discrepancies suggest the surface magnetism of previously studied Cr (001) is different from the $\boldsymbol{Q}_3$ domain (*C*-SDW) here, possibly due to different surface structure or impurity conditions. Before discussing the origin of *C*-SDW, we investigate the boundary structure of different SDW domains first.

### B. The spin/charge structures of *IC*-SDW domain wall

Figures 3(a-c) show the $S_{X,Y,Z}$ maps taken around an in-plane $\boldsymbol{Q}_{1\pm}/\boldsymbol{Q}_{2\pm}$ domain wall. The long period ($\lambda$ = 6.0nm) *IC*-SDW modulation of $\boldsymbol{Q}_{1\pm}/\boldsymbol{Q}_{2\pm}$ are clearly resolved in $S_X/S_Y$ maps, respectively, while $S_Z$ component is absent in the whole region [Fig. 3(c)]. Fig. 3(g) plots the profiles of $S_X/S_Y$ amplitude across the domain wall [by averaging the absolute intensity of Figs. 3(a) and 3(b) along Y direction]. One can see both $\boldsymbol{Q}_{1\pm}/\boldsymbol{Q}_{2\pm}$ decays exponentially at the boundary within a finite range (~10 nm). This behavior reflects the itinerancy of SDW, as the order parameter (spin amplitude) should vary within a scale of coherence length ($\xi$) and does not change abruptly. Therefore, these results give a direct estimation of $\xi$ for *IC*-SDW. From the exponential fittings to $\boldsymbol{Q}_{1\pm}$ and $\boldsymbol{Q}_{2\pm}$ profiles in Fig. 3(g) we obtain $\xi_{//Q}$ = 4.1 (± 0.2) nm and $\xi_{\perp Q}$ = 4.9 (± 0.3) nm (see supplementary material part **IV-1**for fitting details), which characterize the decay of spin amplitude along and perpendicular to *IC*-SDW wavevector, respectively. Such anisotropy of $\xi$ is likely due to anisotropic *IC*-SDW gap opening on Fermi surfaces, as the single-Q nesting breaks the fourfold symmetry of lattice [Fig. 1(b)]. We note that previous neutron scattering study [24] has observed anisotropic magnetic excitation along *Q* and perpendicular to *Q*, which may relate to the anisotropy of $\xi$ here.

Notably, due to the finite decay length, there forms a region where $\boldsymbol{Q}_{1\pm}/\boldsymbol{Q}_{2\pm}$ *IC*-SDW overlap with each other [shaded region in Fig. 3(g)]. This is expected to generate a double-Q *IC*-SDW which does not exist in the bulk. We found such double-Q SDW is further evidenced by the accompanied CDW. Fig. 3(d) shows a non-spin sensitive dI/dV map taken on the same area in Fig. 3(a). Besides the stripe-like CDW associated with $\boldsymbol{Q}_{1\pm}/\boldsymbol{Q}_{2\pm}$ in each domain ($\lambda$ =3.0nm) [36], a square-patterned modulation appears at the boundary (see Fig. S7 for additional data). Fig. 3(e) gives the fast Fourier transform (FFT) of Fig. 3(d). Interestingly, there are four spots locate at (±1/2, ±1/2)$\boldsymbol{Q}_{\mathrm{CDW}}$ (referred as $\boldsymbol{Q}_{DB\text{-}\mathrm{CDW}}$ below) which are not merely the adding of two single-Q CDW. The inversed FFT of $\boldsymbol{Q}_{DB\text{-}\mathrm{CDW}}$ clearly displays a larger squared pattern around domain wall [Fig. 3(f)]. We found these additional CDW can be interpreted as the second order of double-Q SDW. As illustrated Fig. 3(i), the double-Q SDW should result from the coexistence of $\boldsymbol{Q}_{1\pm}$ and $\boldsymbol{Q}_{2\pm}$ at the boundary, hence there will be four folded hole pockets at **Γ** [right part of Fig. 3(i)]. The "second-order" nesting between the folded bands of $\boldsymbol{Q}_{1+}$ and $\boldsymbol{Q}_{2-}$ (and other equivalent nesting) naturally generate all the observed $\boldsymbol{Q}_{DB\text{-}\mathrm{CDW}}$, which cannot form without interference between $\boldsymbol{Q}_{1\pm}$ and $\boldsymbol{Q}_{2\pm}$. This demonstrates that the boundary double-Q SDW is not trivial superposition of single-Q SDWs but a distinct state. In Fig. 3(h) we show the amplitude distribution of $\boldsymbol{Q}_{DB\text{-}\mathrm{CDW}}$ across the boundary (obtained by the spatial lock-in method, see supplementary materials part **VI**). The region where $\boldsymbol{Q}_{DB\text{-}\mathrm{CDW}}$ emerged closely corresponds to the double-Q SDW region in Fig. 3(g), indicating their intimate correlation.

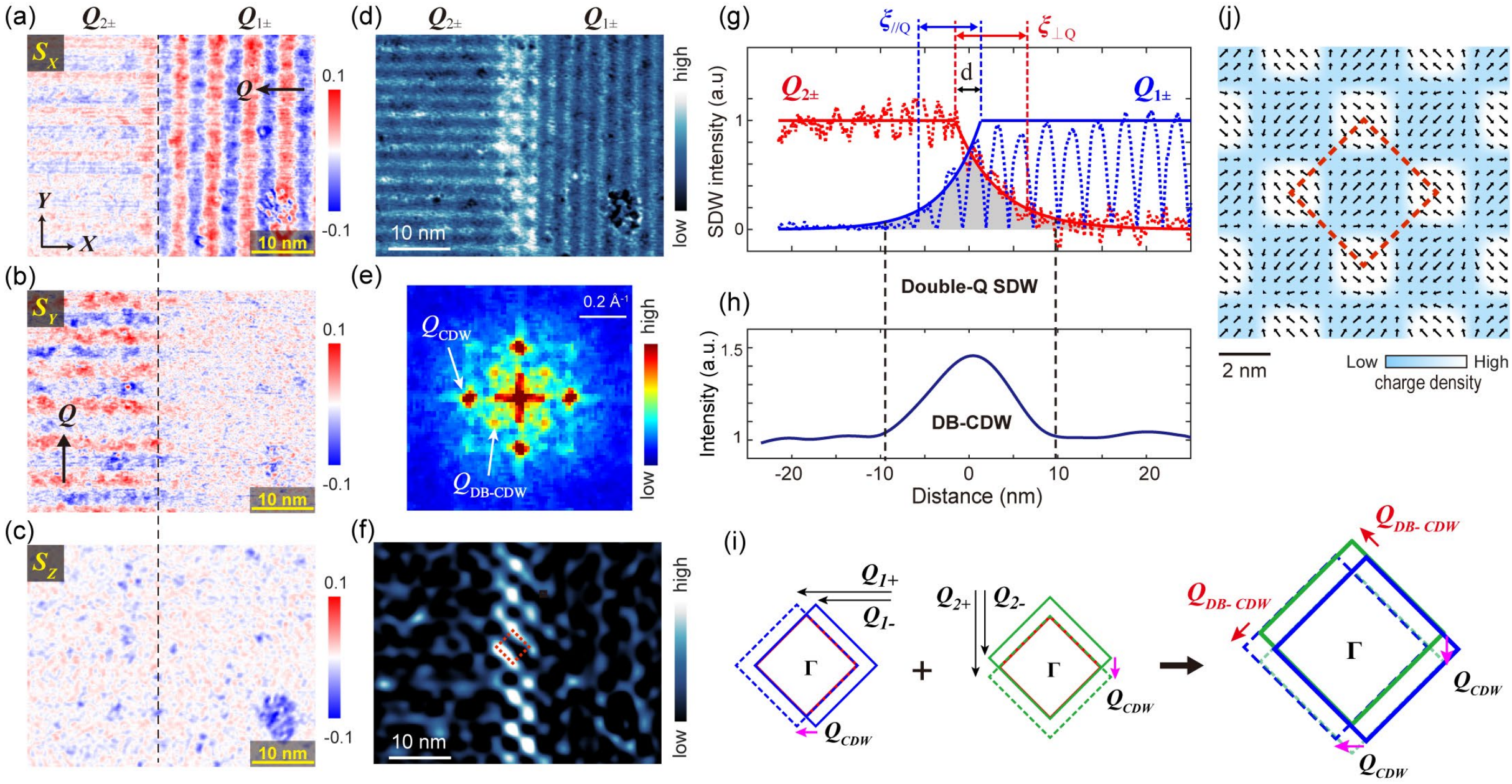

**FIG.3. The in-plane *IC*-SDW domain wall structure.** (a-d) $S_X$, $S_Y$, $S_Z$ and charge maps around *IC*-SDW domain wall [setpoint: $I$ = 50pA, $V_b$ = -100 mV for (a-c), $I$ = 50pA, $V_b$ = -30 mV for (d)]. (e) FFT image of (d), which show the new CDW wavevector of $\boldsymbol{Q}_{DB\text{-}\mathrm{CDW}}$. (f) Inversed Fourier transform of $\boldsymbol{Q}_{DB\text{-}CDW}$ only. (g) Amplitude profile of *IC*-SDW across $\boldsymbol{Q}_{1\pm}/\boldsymbol{Q}_{2\pm}$ domain boundary, by averaging the absolute intensity of panels (a) and (b) along Y direction. (h) Amplitude profile of the $\boldsymbol{Q}_{DB\text{-}\mathrm{CDW}}$ across domain boundary. (i) Illustration of the Fermi surface nesting near the domain boundary, where the co-existence of $\boldsymbol{Q}_{1\pm}$ and $\boldsymbol{Q}_{2\pm}$ generated double-Q *IC*-SDW and the new CDW wavevector $\boldsymbol{Q}_{DB\text{-}\mathrm{CDW}}$. (j) The spin/charge texture of double-Q *IC*-SDW and CDW. Arrows indicate spins and the color indicates charge density. Dashed square is the unit cell of charge modulation, which is also marked in (f).

Therefore, both spin and charge measurements indicate a double-Q *IC*-SDW formed at $\boldsymbol{Q}_{1\pm}/\boldsymbol{Q}_{2\pm}$ domain walls. Fig. 3(j) illustrates the full spin-texture of double-Q *IC*-SDW and the charge modulation of $\boldsymbol{Q}_{DB\text{-}\mathrm{CDW}}$. We note such spin-texture can be viewed as interference of two *IC*-SDW with orthogonal $\boldsymbol{Q}$, much like two itinerant waves meet each other. However, the interference only happens near the domain boundary, as the resulting double-Q *IC*-SDW will have higher energy than single-Q *IC*-SDW [Specifically, for octahedron-shaped electron/hole Fermi surfaces, a single pair of nesting vectors (like $\boldsymbol{Q}_{1\pm}$) already exploits all the Fermi surface phase space, as sketched in Fig. 1(b). Adding $\boldsymbol{Q}$ vectors along other directions does not further gain energy but instead cost it, as discussed in refs.32,33]. It thus represents a new type of magnetic domain wall, distinct in mechanism from the local-moment domain walls where the spins rotate, and can be referred as "interference" wall.

We noticed that a double-Q domain wall was previously observed in a row-wise antiferromagnet (AFM) of monolayer Mn/Re(0001) [44]. Its domain wall energy is dominated by higher-order local exchange interactions. Here the *IC*-SDW in Cr is a collective instability driven by Fermi surface nesting, thus its domain wall inherently requires an itinerant description. As reflected by the shaded region in Fig. 3(g), the width of $\boldsymbol{Q}_{1\pm}$ /$\boldsymbol{Q}_{2\pm}$ interference wall shall be determined by energy difference between double-Q/single-Q *IC*-SDW and the coherence length ($\xi$). We have performed a straightforward calculation (see supplementary material part **IV-2**) to minimize the domain wall energy and find that this relation is given by:

$$\frac{U_{2Q}}{U_{1Q}} = 2 - e^{\frac{d}{2\xi}}$$

Here $U_{2Q}$ and $U_{1Q}$ are the condensation energies (absolute value) of double-Q and single-Q *IC*-SDW, respectively. $d$ is the distance between the points where $\boldsymbol{Q}_{1\pm}$/$\boldsymbol{Q}_{2\pm}$ start to decay [indicated in Fig. 3(g)]. The anisotropy of $\xi$ is neglected here for simplification. From the fitting in Fig. 3(g) we obtain $d$ = 2.9 (±0.3) nm and $\bar{\xi}$=4.5 (±0.2) nm, which yields $U_{2Q}/U_{1Q}$ = 0.62. This is consistent with our expectation that double-Q *IC*-SDW only forms at the domain wall.

For the $\boldsymbol{Q}_{1\pm}$/$\boldsymbol{Q}_3$ domain boundaries [Figs. 2(b,c)], we found the spin amplitude of $\boldsymbol{Q}_{1\pm}$ and $\boldsymbol{Q}_3$ (*C*-SDW) also decays exponentially and the related coherence length is obtained (see supplementary text part **VII**). In this case no additional CDW modulation observed on the boundary (Fig. S3), possibly because the nesting condition of *IC/C*- SDW are different. We note a previous STM study on Cr(110) islands [45] has reported similar domain wall between in-plane/out-of-plane CDW modulation.

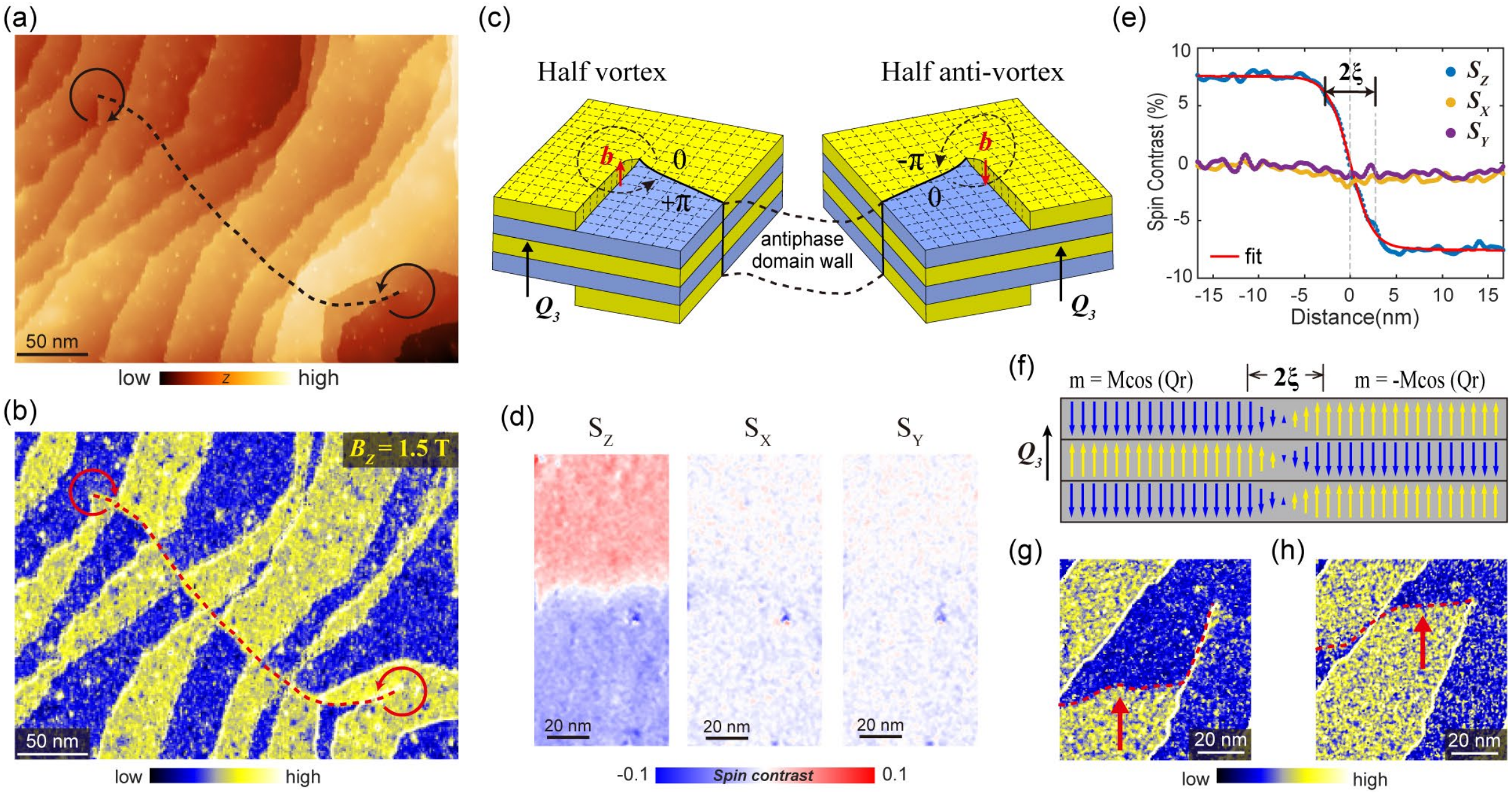

**FIG. 4. Half-vortex and antiphase domain walls in *C*-SDW domain.** (a) Topographic image of a Cr (001) surface with two screw dislocations of opposite chirality. (b) $S_Z$-sensitive dI/dV map of the same region in (a) (Fe tip, $I$ = 100pA, $V_b$ = -120mV, $B_Z$ = 1.5T). An antiphase domain wall (dashed line) connects two screw dislocations. (c) Sketch of spin structure around screw dislocations and antiphase domain wall. The phase of *C*-SDW order parameter changes $+\pi/-\pi$ any loop (dashed curves) encircling screw dislocation. Red arrows represent the Burgers vectors. (d) $S_X$, $S_Y$, $S_Z$ maps around an antiphase domain wall. (e) The line profile of the $S_X$, $S_Y$, $S_Z$ maps. The fitting to $S_Z$ profile yields a decay length ($\xi$) of 2.49 (±0.08) nm. (f) The spin structure of the anti-phase domain wall. (g,h) Spin-resolved dI/dV maps taken in the same area before and after applying tip pulse (Fe tip, $I$ = 100pA, $V_b$ = -250mV), respectively. Red arrows indicate the motion of antiphase domain wall.

### C. The spin/charge structures of C-SDW domain wall and SDW topological defects

Next, we turn to the spin texture and boundary structure observed within $\boldsymbol{Q}_3$ domain. Fig. 4(a) shows a region of this domain with two screw dislocations (which have opposite chirality). Fig. 4(b) is the corresponding dI/dV map with $S_Z$ sensitivity. The *C*-SDW which shows reversed $S_Z$ between adjacent terraces is seen. However, such spin structure gets frustrated when it meets a screw dislocation. As illustrated in Fig. 4(c), around a screw dislocation the (001) plane gradually shifts and merges into its adjacent plane, so the spin orientation must reverse at somewhere in the plane. This will generate an antiphase domain wall of *C*-SDW, as the one tracked by dashed line in Fig. 4(b). In fact, the screw dislocation induces a topological defect of *C*-SDW. That is, after encircling a dislocation core along any loop [Fig. 4(c)], the phase of the order parameter: $\boldsymbol{\eta} = \boldsymbol{m}e^{i(\boldsymbol{Q}_3\cdot\boldsymbol{r}+\varphi_0)}$ will change:

$$\oint d\varphi = \oint \boldsymbol{Q_3} \cdot d\boldsymbol{\mu} = \boldsymbol{Q_3} \cdot \boldsymbol{b} = \pm\pi$$

Here $\boldsymbol{b}$ is the Burgers vector of screw dislocation, which is along out-of-plane direction with $b = a/2 = \lambda_{C\text{-SDW}}/2$. $d\boldsymbol{\mu}$ is the offset of Cr atoms with respected to their original positions. The +/- sign applies for different chirality of screw dislocations. The phase change of $\pm\pi$ (rather than $2\pi$) implies that these topological defects are "half" vortices of SDW. We shall note the half vortices have been generally predicted for inter-layer AFM material with screw dislocations [46,47]. Here the half-integer vorticity is generated by SDW phase change, instead of the rotation of local spin. Interestingly, an antiphase domain wall starts at one screw dislocation will end at another dislocation with opposite chirality, forming a "vortex-antivortex pair". This is clearly observed in Fig. 4(b) and illustrated in Fig. 4(c).

To reveal the spin structure of antiphase domain wall, we present its $S_{X,Y,Z}$ maps in Fig. 4(d), with the corresponding line profiles plotted in Fig. 4(e). The $S_Z$ has dominant weight, which decays to zero at the boundary and reverses the sign. Meanwhile $S_X$, $S_Y$ intensities remain minimal at everywhere. This indicates the total spin density vanished at the antiphase domain wall. Such behavior is in sharp contrast to the Néel/Bloch wall of local-moment magnetism, where the spins rotate to a different direction. The vanishing of spin density is a distinctive hallmark of the itinerant nature of *C*-SDW. As illustrated in Fig. 4(f), the π phase change at the boundary is equivalent to a sign change of order parameter, thus the order parameter (spin amplitude) will have a zero point at the boundary. Formally, it can be treated as a local destructive interference of two antiphase SDW at the boundary, because large-scale suppression of SDW ground state is energetically disfavored. Hence the decay of $S_Z$ towards the boundary should give an estimation of *C*-SDW coherence length, which controls the width of antiphase domain wall. From the exponential fit to $S_Z$ profile in Fig. 4(e) we obtain $\xi_{C\text{-SDW}} \sim 2.49$ (± 0.08) nm. In the BCS mean-field theory, $\xi \sim hv_f/\Delta$, the relatively shorter $\xi$ of *C*-SDW with comparing to *IC*-SDW is likely due to its fully opened gap [Fig. 1(e)].

As seen in Fig. 4(b), the end points of antiphase domain wall are pinned by screw dislocations, however there is no constrains on its specific path. Interestingly, we find the position of domain wall can be moved via STM tip pulse, as illustrated in Figs. 4(g) and 4(h) (before and after pulse, respectively). This demonstrates the possibility of electrical control on SDW domain walls.

We note that antiphase domain wall was also observed in the layer-wise AFM state of previously studied Cr(001) surfaces [39 - 43]. However, the measured domain wall widths in those studies were much larger (~150 – 200 nm) than that of the *C*-SDW domain wall. This discrepancy may arise from differences in magnetic anisotropy induced by varying surface conditions.

### D. SDW in-gap states at the *C*-SDW domain wall

The presence of *C*-SDW gap in $\boldsymbol{Q_3}$ domain prompts further investigation on its response to domain boundary. Figs. 5(a) shows a region containing a screw dislocation, and an antiphase domain wall is seen in its $S_Z$ map [Fig. 5(b)], while Fig. 5(c) is a non-spin sensitive map taken at $V_b$ = 20 mV (inside of *C*-SDW gap). Remarkably, the antiphase wall is also visualized in Fig. 5(c), with locally enhanced tunneling conductance. The dI/dV spectrum taken on the domain wall [Fig. 5(d)] shows a peak feature at ~20 meV with significantly suppressed *C*-SDW gap. This indicates there is confined SDW in-gap state at the antiphase domain wall, as further illustrated in the line-cut spectra taken across the boundary [Fig. 5(e)]. Fig. 5(f) shows the energy dependence of the in-gap states across the boundary. At $E$ = 20 meV, the in-gap state is localized within a width (FWHM) of ~ 7 nm. As energies goes higher, the distribution gradually expands and disappears at E> 50 meV (corresponds to the SDW gap edge).

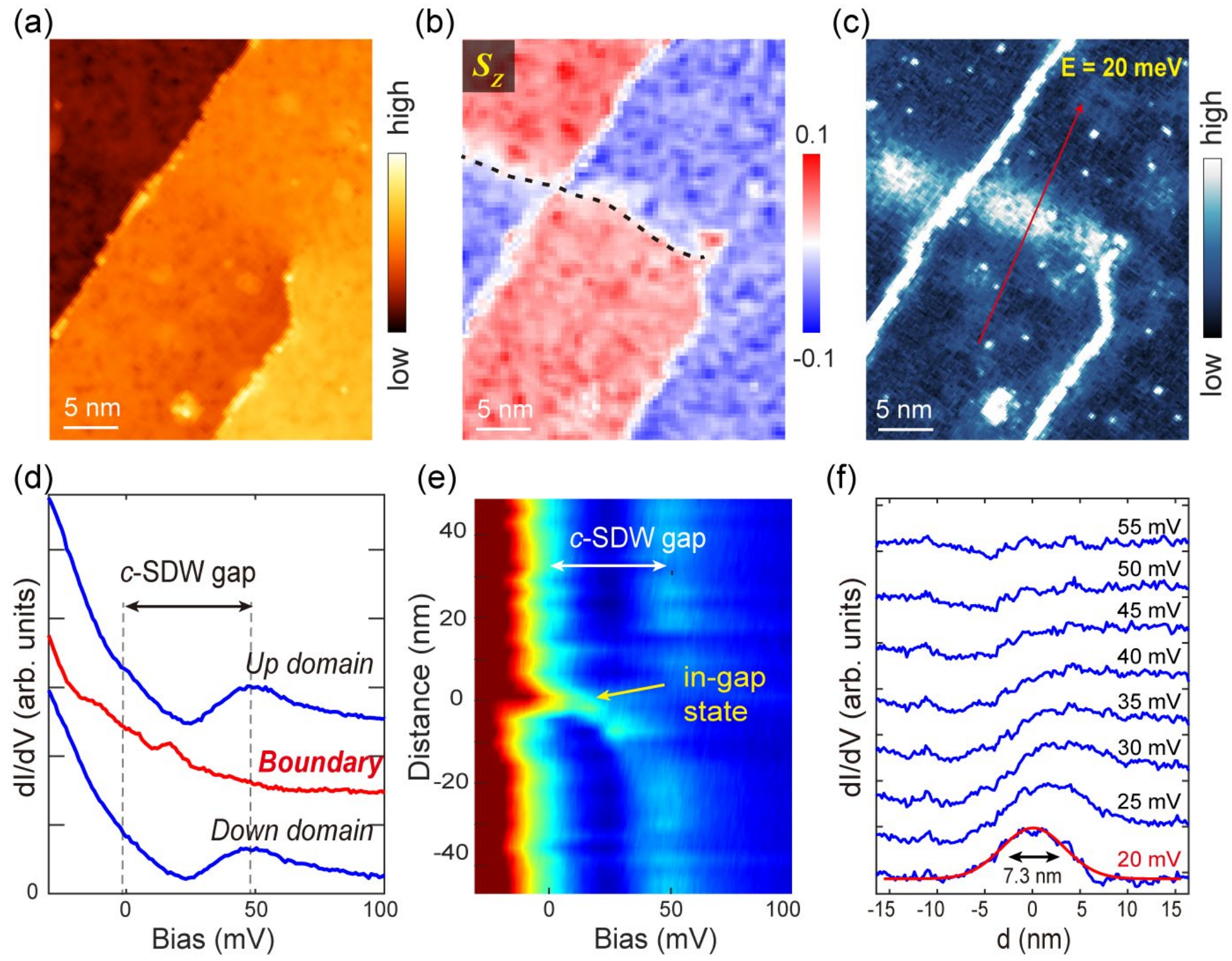

**FIG.5. SDW quasiparticle state at the antiphase domain wall. a-c**, Topographic image, Sz map and non-spin sensitive dI/dV map ($V_b$ = 20 mV) taken around a screw dislocation. The antiphase domain wall is visualized in (**c**) which terminated at the screw dislocation. **d**, dI/dV spectra taken on and off the antiphase domain wall ($I$ = 100 pA, $V_b$ = 100mV) **e**, Color plot of dI/dV spectra taken across the antiphase domain wall. **f**, Line profiles of the dI/dV map across an antiphase domain wall, showing the spatial distribution of in-gap state at different energies.

## III. THEORETICAL MODELING OF SDW BOUNDARY STATE

Under the framework of mean-field description, the closing of SDW gap at antiphase domain wall coincides with the vanishing of SDW order parameter and spin amplitude. This is analogous to an interface where a superconducting gap changes sign or a phase of π [8,48], and where Andreev bound state will emerge. Consequently, we expect the observed in-gap states at the antiphase wall are SDW boundary state. To illustrate it theoretically, we performed a model calculation on the electronic state of SDW antiphase walls. The calculation employed a Fermi surface similar to that shown in Fig. 1(b). We use the annihilation field operators $\psi_{\Gamma,X,Y}$ to describe single particle states near the associated Fermi pockets at $\boldsymbol{\Gamma}, \boldsymbol{X}, \boldsymbol{Y}$, and we shift the center of mass of the pockets so that all of them are around the zone center. The SDW order will then scatter electrons from $\boldsymbol{\Gamma}$ to $\boldsymbol{X}$ $(\boldsymbol{Y})$ and vice versa. The effective Hamiltonian for a uniform SDW system can be written as, in the momentum space:

$$H = \sum_{k,\alpha=\Gamma,X,Y} \psi^{+}_{\alpha,k} \epsilon^{\alpha}_{k} \psi_{\alpha,k} - \sum_{k,\alpha=X,Y} \left( \psi^{+}_{\Gamma,k} m_{\alpha} \sigma_{\alpha} \psi_{\alpha,k} + h.c. \right).$$

Here the spin indices are left implicit, $\epsilon^{\alpha}_{k}$ is the energy dispersion for the $\alpha$-pocket, $\sigma_{\alpha}$ is the Pauli matrix in the spin basis, and $m_{\alpha}$ is the spin scattering potential from the SDW order parameter. In the uniform SDW state, $m_{\alpha}$ is a constant, while in the antiphase domain wall configuration, we assume it changes as $m_\alpha(x) = m_0 \tanh(x/\xi)$, where $m_0$ is the global amplitude, $x$ is the displacement normal to the domain wall, and $\xi$ is the SDW coherence length (more details of the model are described in Supplementary Materials part **VIII**). We find that if the single-Q SDW wavevector is along X, the Y pocket is an idle pocket (no SDW gap opening), and vice versa. Moreover, for the two pockets scattered by the SDW order, one of them is electron-like and the other hole-like, so the spin scattering term is similar to the local pairing term in the Bogoliubov-de Genes Hamiltonian of superconductivity. Therefore, the antiphase domain wall is mapped to a Josephson junction with a phase shift of π. This gives rise to an "Andreev bound state" of SDW, as our numerical calculations show below.

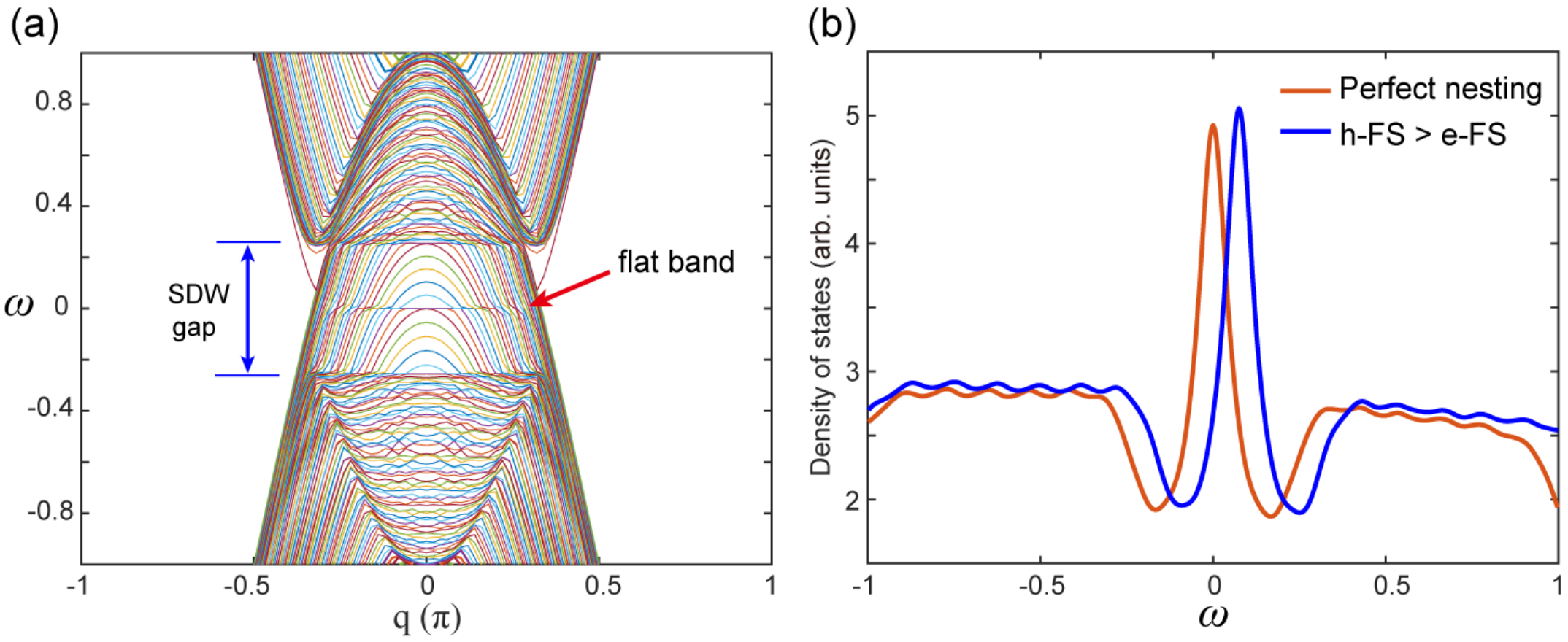

**FIG. 6.** (a) Energy dispersion versus the momentum along antiphase domain wall. (b) The calculated local density-of-state (DOS) at antiphase domain wall, with perfect nesting condition (red) and the case of hole pocket larger that electron pocket (bule). (DOS is summed over a few sites near the center of the domain wall).

Figure 6(a) shows the calculated energy as a function of the conserved momentum along the antiphase domain wall, for the case of perfect Fermi surface nesting (the hole and electron pockets have the same size). A flat band is clearly observed inside the SDW gap, which is exactly derived from the in-gap bound state. The other dispersive bands in SDW gap are from ungapped hole pocket. Fig. 6(b) shows the calculated DOS at the domain wall. For perfect Fermi surface nesting, there is a zero-bias peak at the gap center; while in the case of non-perfect nesting (the hole pocket is larger than electron-pocket), the gap is shifted to positive energy and so does the bound state energy. This is exactly what we observed in Fig. 5(d). We note that unlike the superconductivity which happens in particle-particle channel, SDW does not ensure particle-hole symmetry and thus the gap can open above/below $E_F$, leading to the boundary states off zero-bias.

## IV. DISCUSSION AND CONCLUSION

We have now presented a comprehensive study on the boundary structure of SDW in Cr. We unveiled various feature that originate from the itinerancy of SDW, including the finite scale decay of spin amplitude, the double-Q SDW/CDW, the half-vortex and antiphase domain walls with vanished spin density. Both in-plane *IC*-SDW domain wall and *C*-SDW antiphase domain wall can be treated as local interference of two SDW, either with different ***Q*** or reversed phases. Therefore, they represent a new magnetic domain wall structure that distinct from those of local-moment magnets [11,49], named as interference wall here. We show that the energy and width of these interference walls are determined by interference induced double-Q SDW and/or SDW coherence length, rather than the local exchange interaction and magnetic anisotropy which dominate conventional magnetic domain walls. The possibility of moving SDW domain wall by electric pulse is also demonstrated. These results could pave a way of utilizing SDW material for spintronic applications.

The observation of in-gap quasiparticle states at *C*-SDW antiphase domain wall is another key finding of this study. To our knowledge, it is the first time the quasiparticle bound state of SDW has been visualized. Such state can be described by the similar mean-field Bogoliubov-de Gennes (BdG) equation of superconductivity, as both superconductivity and SDW are itinerant quantum orders. Unlike the Bogoliubov quasiparticles which are mixtures of particle and holes [2,3], the SDW quasiparticles are mixtures of spin-up and spin-down particles [2]. How this unique property manifest itself is worthy of further studies. We note some theoretically works suggesting SDW quasiparticles may coexist and even drive unconventional superconductivity in correlated materials [50,51].

Furthermore, our work also provides insights on the mechanism of SDW and its interplay with CDW. The structure of in-plane *IC*-SDW in Cr is well accounted by the "imperfect" Fermi surface nesting model. However, an *IC*-SDW to *C*-SDW transition happens when ***Q*** is along out-of-plane. This is likely due to the surface breaks the translation symmetry for out-of-plane SDW, which affects the nesting condition and causes significant variation of long wavelength *IC*-SDW [52,53]. The possible enhanced magnetic moments predicted at Cr (001) surface may be another reason to force the SDW became commensurate [54]. Nonetheless, the gap opening in the electronic density-of-states and non-rotation spin at antiphase domain wall still suggest a dominant itinerant nature of *C*-SDW. The observation of a new CDW modulation at *IC*-SDW domain wall is consistent with previous theoretical works suggesting CDW as high order harmonics of SDW [30,31].

Overall, our observation of interference domain wall of *IC*/*C*-SDW and SDW quasiparticle states provide new insights on the comprehensive understanding of itinerant magnetism. We expect such interference domain wall could widely exist in other forms of itinerant magnets, which can drive more exotic spin/charge structures that do not exist in the bulk. Our study will inspire further research on itinerant magnetic domain walls and prompt their applications in future's spintronics.

## ACKNOWLEDGMENTS

We thank professors X.G. Gong, Y. Qi, J. Xiao, J.X. Li, X.C. Xie, and Y.Y. Wang for helpful discussion. The work is supported by National Natural Science Foundation of China (Grants Nos.:12225403, 92365302, 92365203, 12374147, 12304181, 12104094), Innovation Program for Quantum Science and Technology (Grant no.: 2021ZD0302803), The New Cornerstone Science Foundation (China), Shanghai Municipal Science and Technology Major

Project (Grant No. 2019SHZDZX01), Shanghai Pilot Program for Basic Research (Grant No. 21TQ1400100).

**Data Availability:**
All raw data corresponding to the finding in this manuscript are available at:
**https://doi.org/10.6084/m9.figshare.29588708**

## APPENDIX: METHODS

**Cr sample preparation:** The Cr (001) single crystal (Mateck, 99.999% purity) with a size of 4mm×4mm was mounted on a sample holder by gently clamping its four corners (no glue). To obtain clean surface, the sample was treated by repeated cycles of Ar sputtering at 750°C (for 15 min) and annealing at 800°C (for 20 min). Cleaned sample was transferred to STM chamber under ultra-high vacuum (no magnetic field applied during the cooling).

**SP-STM measurements:** The spin-polarized STM experiment was conducted in a cryogenic STM system (UNISOKU) at T = 4.5 K. Spin-resolved tunneling spectra and conductance mapping were measured by Fe coated tips, which were prepared by depositing 10~20 nm thick Fe film on W tip. The W tip was electrochemically etched and flashed up to ≈ 2000K for cleaning before coating. The tunneling conductance (dI/dV) was collected by standard lock-in method and the bias voltage ($V_b$) is applied to the sample.

# Supplementary Materials for

## Discovery of Itinerant Magnetic Domain Wall and Quasiparticle Boundary State in Spin-Density-Waves

Yining Hu†, Xu Wang†, Chen Chen*, Qingle Zhang, Dongming Zhao, Tianzhen Zhang, Chenxi Wang, Qiang-Hua Wang, Donglai Feng*, Tong Zhang*

†These authors contributed equally to this work.
*Corresponding authors.
Email: cchen_physics@zjnu.edu.cn, dlfeng@ustc.edu.cn, tzhang18@fudan.edu.cn

### Part- I. Measurement of pure spin-contrast along X, Y, Z directions.

Using Fe-coated tip, we can measure dI/dV maps with spin sensitivity along X ([100]), Y ([010]) and Z ([001]) directions, which is achieved by polarizing the tip under a vector magnetic field along X, Y and Z, respectively. The pure spin contrast maps ($S_X$, $S_Y$, $S_Z$) can be obtained by the relative difference of dI/dV maps taken under $\pm B_X$, $\pm B_Y$ and $\pm B_Z$, respectively. For example, Figs. S1(a) and S1(b) show two dI/dV maps taken under $B_X$ = 1T and $B_X$ = -1T around $Q_{1\pm}/Q_{2\pm}$ domain wall. The $S_X$ map is obtained by $\frac{dI/dV_{B_x} - dI/dV_{B_{-x}}}{dI/dV_{B_x} + dI/dV_{B_{-x}}}$, which gives pure spin contrast along X direction ($S_X$ map). Similarly, Figs. S1(d,e) show the dI/dV maps taken under $\pm B_Z$ around a $Q_{1\pm}/Q_3$ domain wall, and Fig. S1(f) is the corresponding $S_Z$ map. We note that the spin sensitivity of Fe tip is energy (bias voltage) dependent, the spin-resolved maps were usually taken with $V_b$ = -100 mV to -180 mV, which has relatively high spin sensitivity. We shall note that since the SDW of Cr (an AFM state in general) is quite robust against magnetic field (ref. 19), the applied field here (|B|<1.5T) does not affect the spin structure of Cr but only polarize the Fe tip.

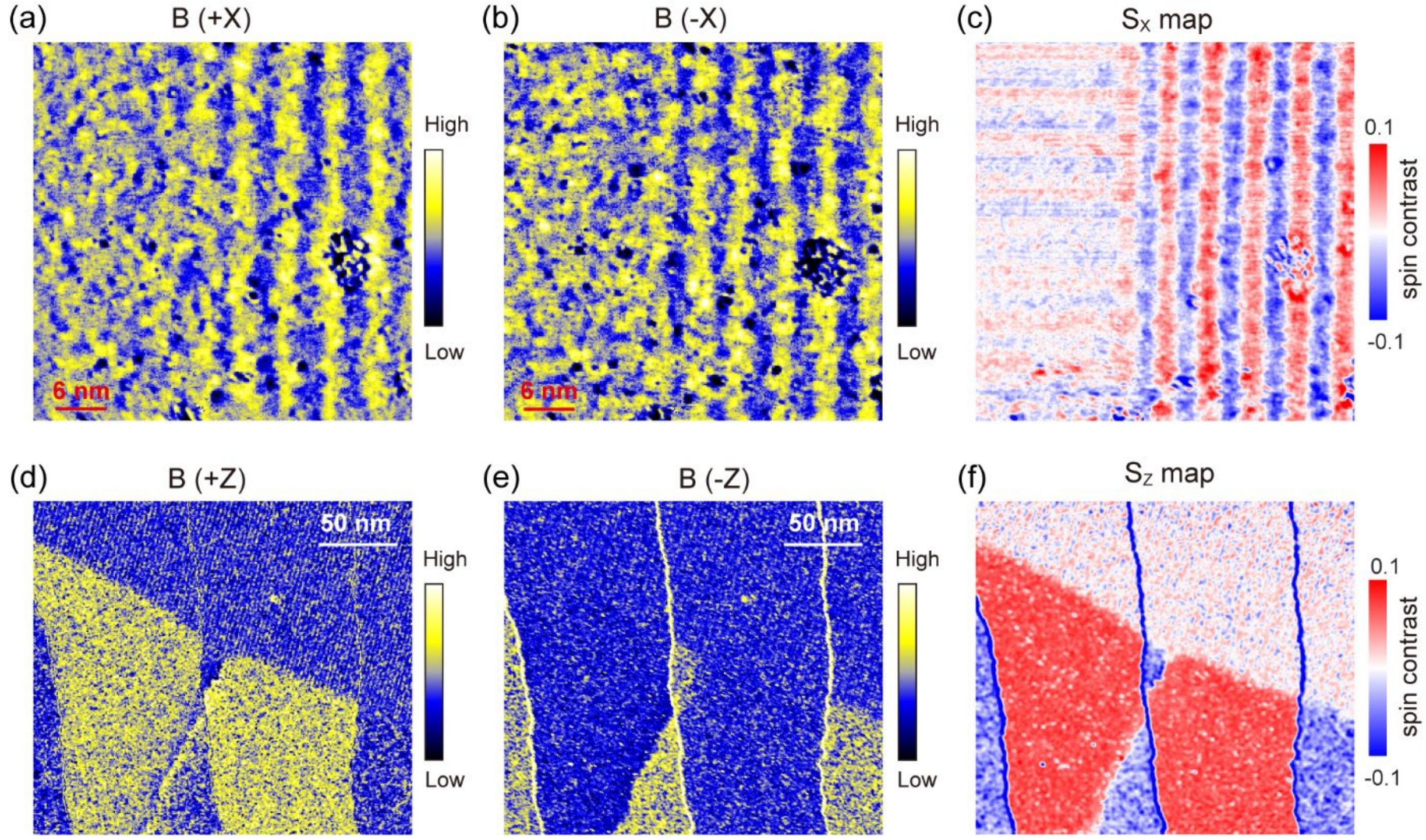

**Fig. S1.** (a,b) dI/dV maps taken in the same area by Fe-coated tip under Bx = 1T and -1T, respectively (setpoint: $I$ = 50 pA, $V_b$ = -100 mV). (c) Pure spin-contrast map (**$S_X$** map) obtained by calculating the relative difference of panel (a) and (b). (d,e) dI/dV maps taken in the same area by Fe-coated tip under Bz = 1.5T and -1T, respectively (setpoint: $I$ = 50pA, $V_b$ = -120mV). (f) Pure spin-contrast map (**$S_z$** map) obtained by calculating the relative difference of dI/dV maps in (d) and (e).

## Part- II. Large scale spin maps of $Q_3$ domain.

Figs. S2(a,b) show the large scale topographic image and spin-resolved dI/dV map in $Q_3$ domain. The spin contrast is along $Z$ direction, and panel b shows the adjacent terraces have opposite $S_Z$ component. Fig. S2(c) shows the line profile along the arrow in panel (b). We can see the spin contrast almost remains constant even over 17 terraces (~2.4nm), which means the long-period *IC*-SDW modulation is absent and evidences the SDW became commensurate in $Q_3$ domain.

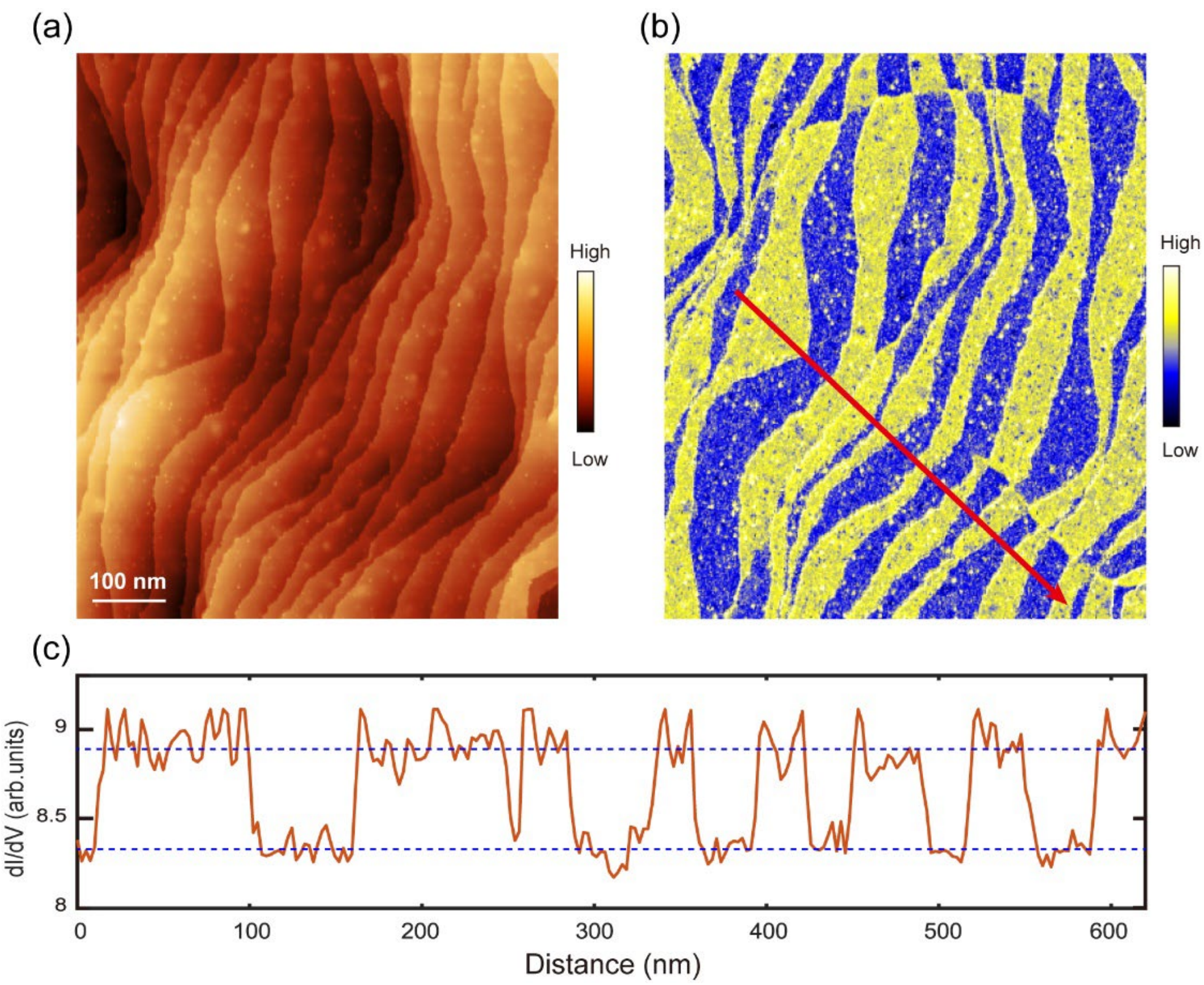

**Fig. S2.** (a) Topographic image of Cr(001) surface in $Q_3$ domain (setpoint: $I$ = 100pA, $V_b$ = -120mV). (b) Spin-resolved dI/dV map of the same region in (a), (Fe-coated tip, Bz = 1.5T). (c) The line profile along the arrow in panel b, shows no long-period modulation over a series of terraces.

## Part- III. Large energy scale dI/dV spectra and electronic structure of $\boldsymbol{Q}_{1\pm}$ and $\boldsymbol{Q}_3$ domains

Fig. S3(a) shows the spin-insensitive dI/dV map taken around $\boldsymbol{Q}_{1\pm}$/$\boldsymbol{Q}_3$ domain wall. In $\boldsymbol{Q}_{1\pm}$ domain, we can see clear in-plane incommensurate CDW with a period of 3.0 nm, and in $\boldsymbol{Q}_3$ domain no *IC*-CDW observed. Fig. S3(b) shows the large energy scale dI/dV spectra taken on both domains. These spectra have very similar line shape with a pronounced DOS peak located at ≈ -70 mV, which is most likely from the surface state of Cr (ref. 21). The difference is that a *C*-SDW gap can be identified in $\boldsymbol{Q}_3$ domain (indicated by black arrows) while it is absent $\boldsymbol{Q}_{1\pm}$ domain. The dI/dV spectra taken across the domain wall is shown in Fig. S3(c). In the $\boldsymbol{Q}_{1\pm}$ domain, there is a dip feature near -22 mV in dI/dV which is related to the incommensurate CDW; while in the $\boldsymbol{Q}_3$ domain the dip feature disappeared and a gap feature related to *C*-SDW appears between 0-50 mV, which is interpreted as the *C*-SDW gap.

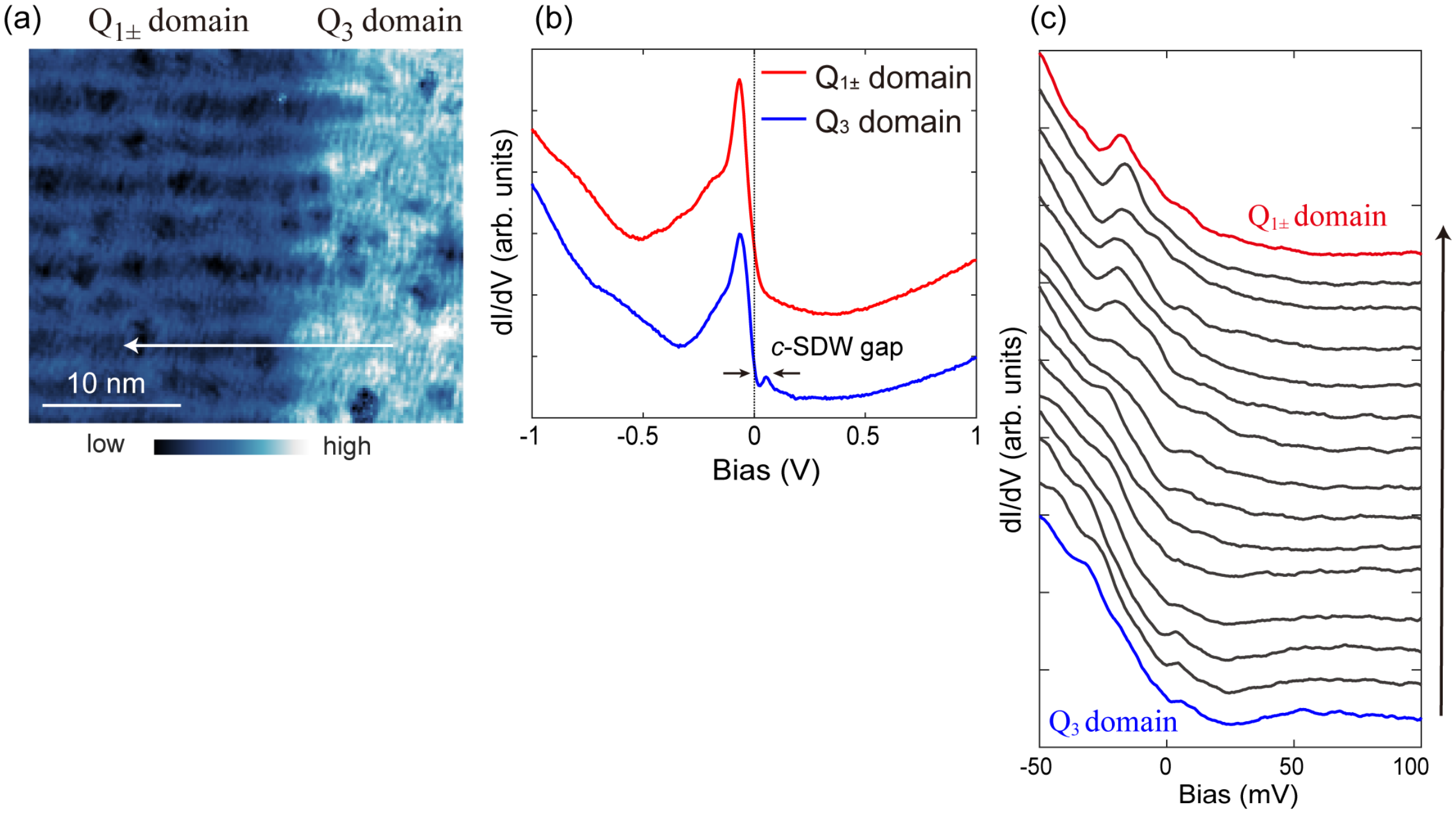

**Fig. S3.** (a) Spin-insensitive dI/dV map taken around $\boldsymbol{Q}_{1\pm}/\boldsymbol{Q}_3$ domain wall ($I$ = 50pA, $V_b$ = -30 mV). (b) Large energy scale dI/dV spectra taken on $\boldsymbol{Q}_{1\pm}$ and $\boldsymbol{Q}_3$ domains (Setpoint: $I$ = 60 pA, $V_b$ = 1 V for $Q_{1\pm}$ domain; $I$ = 200 pA, $V_b$ = -1 V for $Q_3$ domain). (c) A series of dI/dV spectra taken along the arrow shown in panel (a) (setpoint: $I$ = 40pA, $V_b$ = -50 mV).

## Part- IV. The spin/charge structure, energy, and width of double-Q *IC*-SDW domain wall

### IV-1: The spin/charge structure

The spin structure of double-Q SDW domain wall is obtained by fitting the profile of $S_X$, $S_Y$ maps [Figs. 3(a,b)]. Below we show the fitting details. The profile of each *IC*-SDW domain are fitted by a piecewise function. For the domain on left side the SDW amplitude is directly measured, and the fitting function is:

$$f(x) = \begin{cases} 1 , & 0 \le x \le a \\ \exp\left(\dfrac{-x+a}{\xi}\right) , & x > a \end{cases}$$

Here $a$ is position where the SDW amplitude starts to decay. $\xi$ is the characteristic length of the exponential decay. The fitting is performed by using standard Matlab fitting tools, and the results are shown in Fig. S4. The fitting errors of $a$ and $\xi$ can be read in the legend.

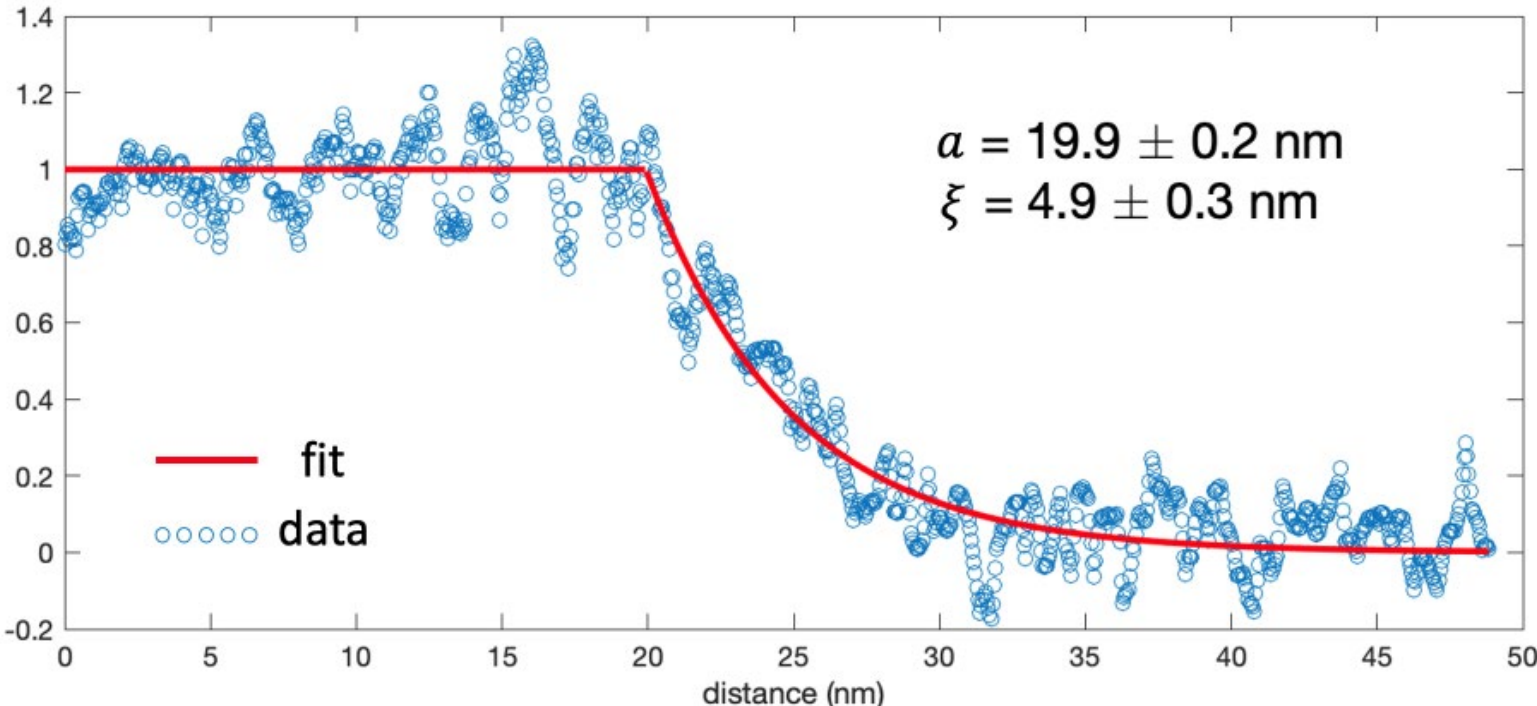

**Fig. S4.** The fitting of the averaged $S_Y$ intensity [absolute value of Fig. 3(b)] along x direction.

For the domain on the right side [Fig. 3(a)], the line profile shows decayed SDW modulation, and the fitting function is:

$$f(x) = \begin{cases} \exp\left(\frac{x-a}{\xi}\right) \cdot \cos\left(\frac{2\pi x}{\lambda} - \varphi_0\right), & 0 \le x \le a \\ 1 \cdot \cos\left(\frac{2\pi x}{\lambda} - \varphi_0\right), & x > a \end{cases}$$

Similarly, the free parameters are $a$ and $\xi$. $a$ is the position where SDW modulation starts to decay. $\lambda$ is the wavelength of *IC*-SDW which is measured to be 5.8 nm, and the value of $\varphi_0$ is set as 2.0 rad. The fitting results are shown in Fig. S5. The fitting errors can be read in the legend.

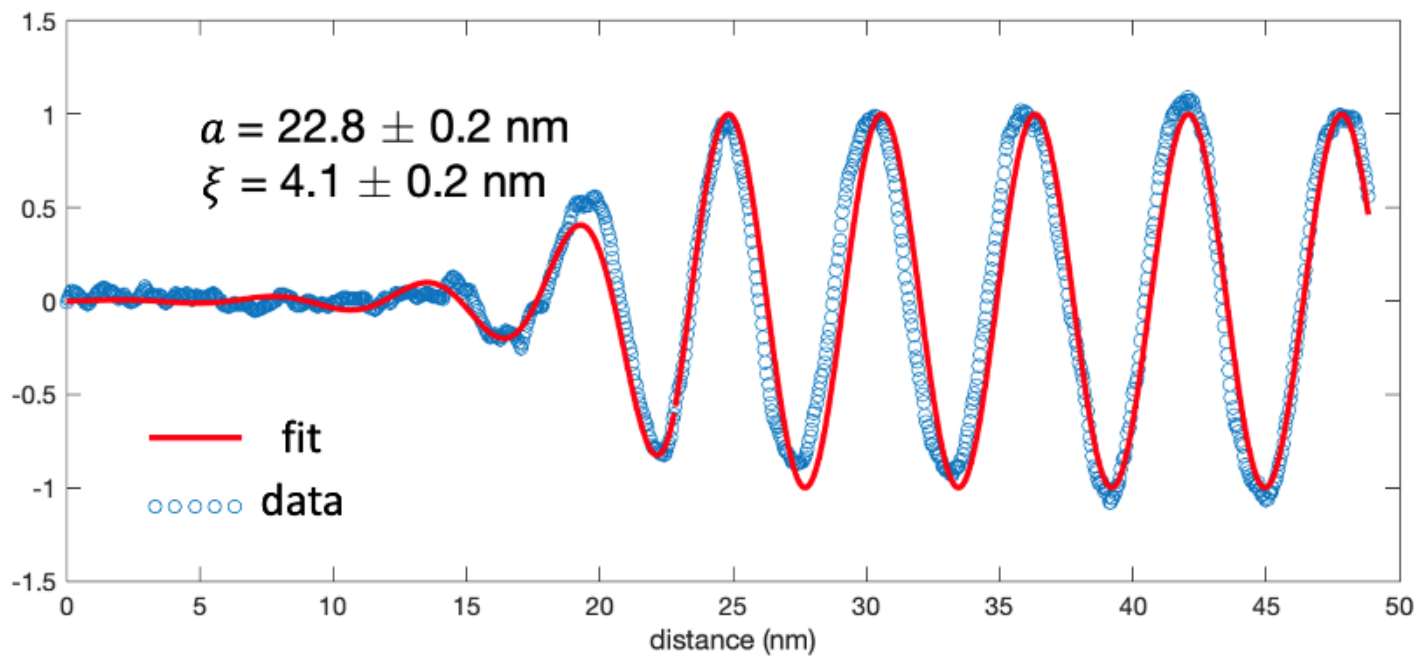

**Fig. S5.** The fitting of the $S_X$ intensity [Fig. 3(a)] along x direction.

Then, we put the fitted parameters into the function describing the intensity of I*C*-SDW amplitude:

$$f(x) = \begin{cases} \exp\left(\frac{x-a}{\xi}\right), & 0 \le x \le a \\ 1, & x > a \end{cases}$$

This gives the solid blue curves in Fig. 3(g).

### IV-2: The energy/width of double-Q *IC*-SDW domain wall

The domain wall energy and width of $\boldsymbol{Q}_{1\pm}/\boldsymbol{Q}_{2\pm}$ *IC*-SDW should be determined by the condensation energy difference between single-Q and double-Q *IC*-SDW and the coherence length. Fig. S6 illustrates such domain wall structure along a line profile perpendicular to the wall. The *IC*-SDW amplitudes ($A_{SDW}$) in the $\boldsymbol{Q}_{1\pm}/\boldsymbol{Q}_{2\pm}$ domains are normalized to 1 and they decay exponentially near the boundary. $d$ is the distance between the two points where $\boldsymbol{Q}_{1\pm}$ and $\boldsymbol{Q}_{2\pm}$ start to decay. To simplify the calculation, we let $\boldsymbol{Q}_{1\pm}$ and $\boldsymbol{Q}_{2\pm}$ have the same decay length of $\xi$ = 4.5 nm (obtained from the fitting in Fig. 3(g)), and assume the energies of single/double-Q SDW have a linear dependence with the area they occupied in Fig. S6. We define the energy without any SDW order is zero ($U_{\text{vac}}$ = 0). The condensation energy of single-Q SDW is -$U_{1\text{Q}}$ (per area in Fig. S6), and that of double-Q SDW is -$U_{2\text{Q}}$. Then the total energy of the system containing an *IC*-SDW domain wall is expressed as:

$$U = \int_{-\infty}^{\infty} -U_{1Q}\, A_{1Q}(x)dx + \int_{-\infty}^{\infty} -U_{2Q}A_{2Q}(x)dx$$

$$= -2U_{1Q}\left(\int_{-\infty}^{-\frac{d}{2}}\left(1 - e^{\frac{x-\frac{d}{2}}{\xi}}\right)dx + \int_{-\frac{d}{2}}^{0}\left(e^{-\frac{x+\frac{d}{2}}{\xi}} - e^{\frac{x-\frac{d}{2}}{\xi}}\right)dx\right) - 2U_{2Q}\int_{-\infty}^{0} e^{\frac{x-\frac{d}{2}}{\xi}}dx$$

The distance of $d$ reflects the domain wall width, which also determines the domain wall energy. If $d$ is much larger than $\xi$, the decrease of $d$ will lower system's energy by forming single-Q SDW. However, the double-Q SDW region increases as $d$ decreases, which will raise the energy. Therefore, at certain $d$ the total energy shall reach a minimum. From the above expression, when $U$ reaches its minimum, the $d$ value satisfies:

$$\frac{U_{2Q}}{U_{1Q}} = 2 - e^{\frac{d}{2\xi}}, \qquad d \in [0, \infty)$$

By taking the measured value of $d$ = 2.9 (±0.3) nm and $\xi$ = 4.5 (±0.2) nm, we get $U_{2Q} = 0.62\, U_{1Q}$. This agrees with our assumption that single-Q state has larger condensation energy.

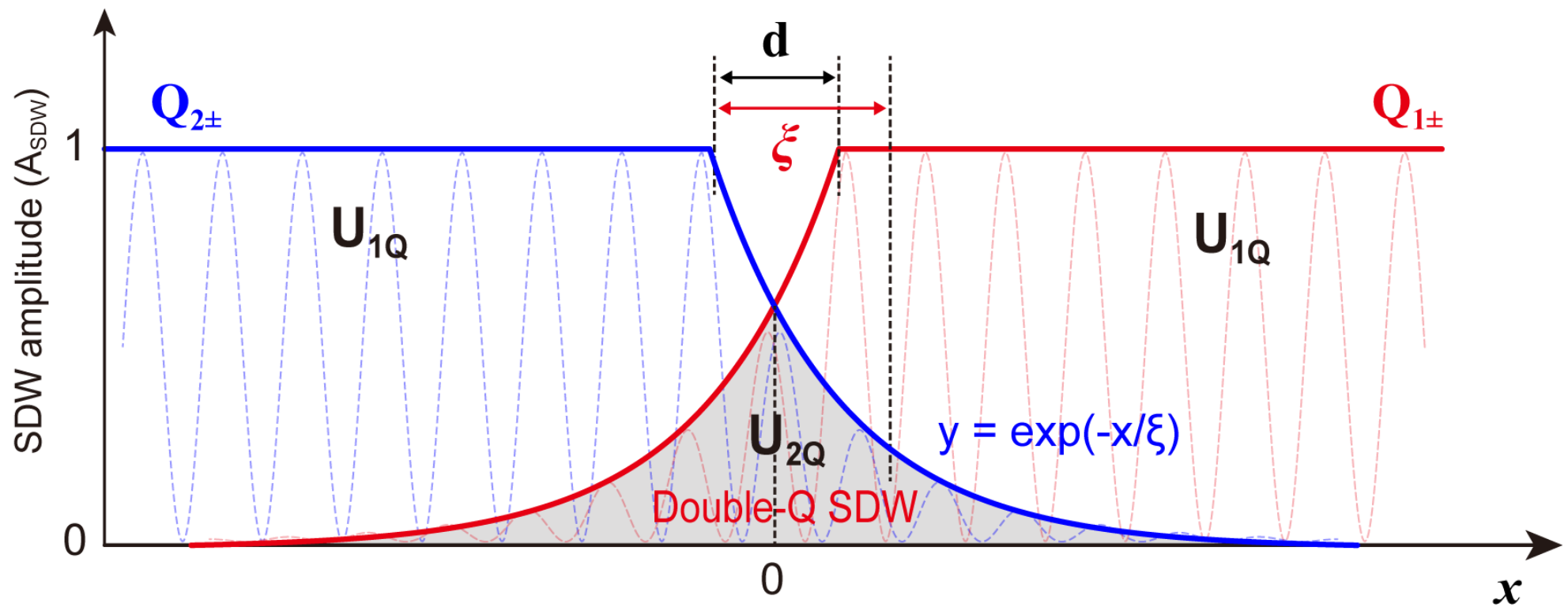

**Fig. S6.** Illustration of in-plane IC-SDW domain wall. The region with single-Q SDW has condensation energy of $U_{1Q}$ (per length along $\boldsymbol{x}$); while the double-Q SDW region (shaded region) has a condensation energy of $U_{2Q}$ (per length along $\boldsymbol{x}$).

## Part- V. Additional data about the CDW modulation on the *IC*-SDW domain wall.

In Fig. S7 we show additional CDW maps taken around *IC*-SDW domains wall and their FFT images. The $\boldsymbol{Q}_{\text{DB-CDW}}$ modulation are clearly seen in these maps (its unit cell is indicated by the dashed square). Figs. S7(a,b) are taken by W tip with magnetic coating, but under spin-insensitive bias voltage.

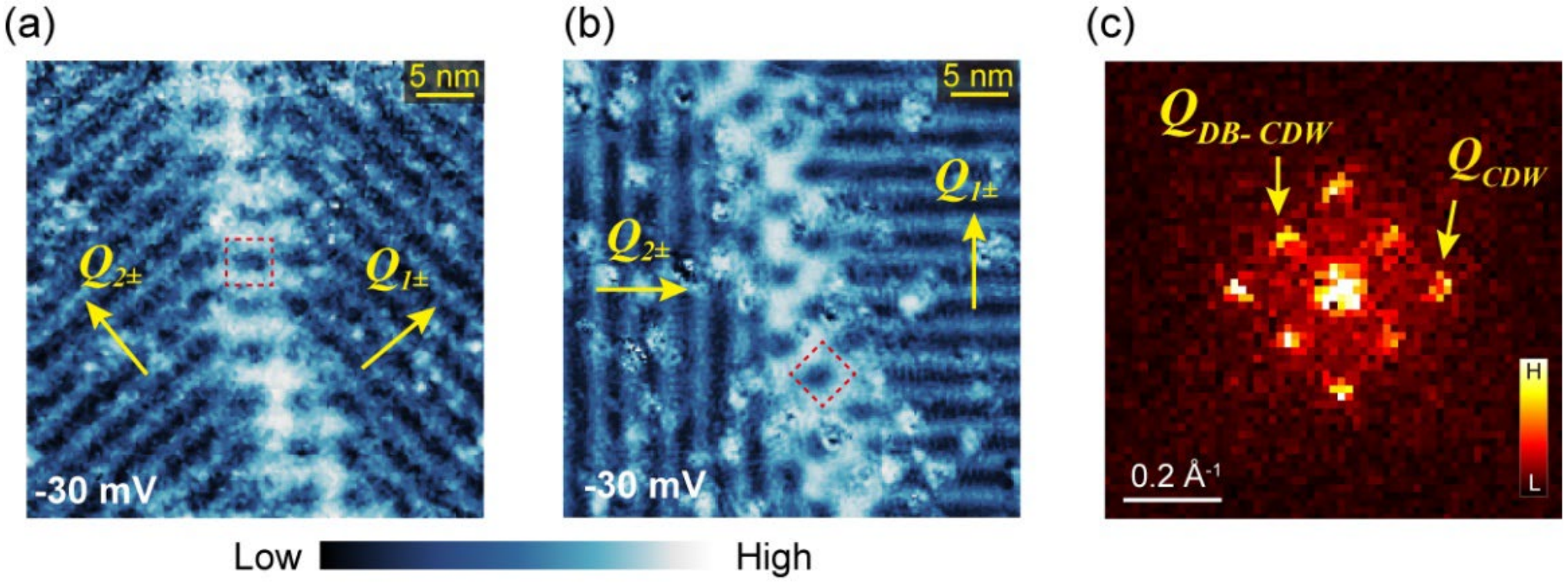

**Fig. S7** (a,b) spin-insensitive dI/dV maps taken near *IC*-SDW domain walls with different orientation, which all show the charge modulation of $\boldsymbol{Q}_{\text{DB-CDW}}$. (Setpoint: $I$ = 50 pA, $V_b$ =-30 mV). (c) FFT image of panel (a).

## Part- VI. The spatial lock-in method for extracting the amplitude of $\boldsymbol{Q}_{\text{DB-CDW}}$ modulation across the boundary.

The spatial lock-in method is a technique which can visualize the amplitude of a specific modulation in real space (refs. S1-S4). To do that, we first generate a 2D Gaussian function $f(x,y) = \exp(-\frac{(x-x_0)^2+(y-y_0)^2}{2\sigma^2})$ which reaches its maxima at the center position $(x_0, y_0)$ and decreases isotropically away from the center with the broadening of Gaussian peak determined by $\sigma$. Then, we do element-by-element multiplication of this 2D Gaussian matrix $f(x,y)$ and the data matrix $R(x,y)$ we want to analyze. This will return a matrix $R'(x,y) = f(x,y).* R(x,y)$, whose data information is majorly preserved around position $(x_0, y_0)$ but mostly lost far away from position $(x_0, y_0)$. After applying 2D FFT transformation to the returned matrix $R'(x,y)$, we can extract the amplitude $A(x_0, y_0)$ of the order with wave vector $\boldsymbol{Q}$ from the absolute value at the corresponding $\boldsymbol{Q}$ position in FFT. Here, since we want to extract the amplitude information for an order with bidirectional modulation, we take an average for the absolute value of two $\boldsymbol{Q}_{DB-CDW}$ who are perpendicular to each other. If we do the above analysis for each pixel of $R(x,y)$ by changing the parameter $x_0$ and $y_0$ in $f(x,y)$, then we can obtain the amplitude distribution map $A(x,y)$ of the $\boldsymbol{Q}$ order. By averaging elements in matrix $A(x,y)$ along the wall direction, we obtained the amplitude distribution of $\boldsymbol{Q}_{DB-CDW}$ across the domain boundary as shown in Fig. 3(h). Parameter $\sigma$ we used to obtain Fig. 3(h) is 3.9 nm.

## VII. The spin structure of $\boldsymbol{Q}_{1\pm}/\boldsymbol{Q}_3$ domain wall:

Fig. S8 shows the spin structure of the $\boldsymbol{Q}_{1\pm}/\boldsymbol{Q}_3$ domain wall. The spin contrast of $\boldsymbol{Q}_{1\pm}$ and $\boldsymbol{Q}_3$ can be (and only be) visualized in $S_X$ and $S_Z$ maps, respectively. The amplitude of $\boldsymbol{Q}_{1\pm}$ (*IC*-SDW) and $\boldsymbol{Q}_3$ (*C*-SDW) both decays exponentially at the boundary. Exponential fits to the decay yield $\xi_{\perp Q1}$ = 3.62 (±0.10) nm and $\xi_{\perp Q3}$ = 3.28 (±0.10) nm.

We note that in ref. 45 of the main text, P.-J- Hsu et al. investigated the CDW modulation in wedge-shaped Cr(110) islands. At certain thickness, the q-vector of CDW shifted to out-of-plane direction, resulting in a boundary between in-plane (projected) and out-of-plane CDW domains. This domain wall displays similar appearance with the $\boldsymbol{Q}_{1\pm}/\boldsymbol{Q}_3$ domain wall shown in Fig. S8.

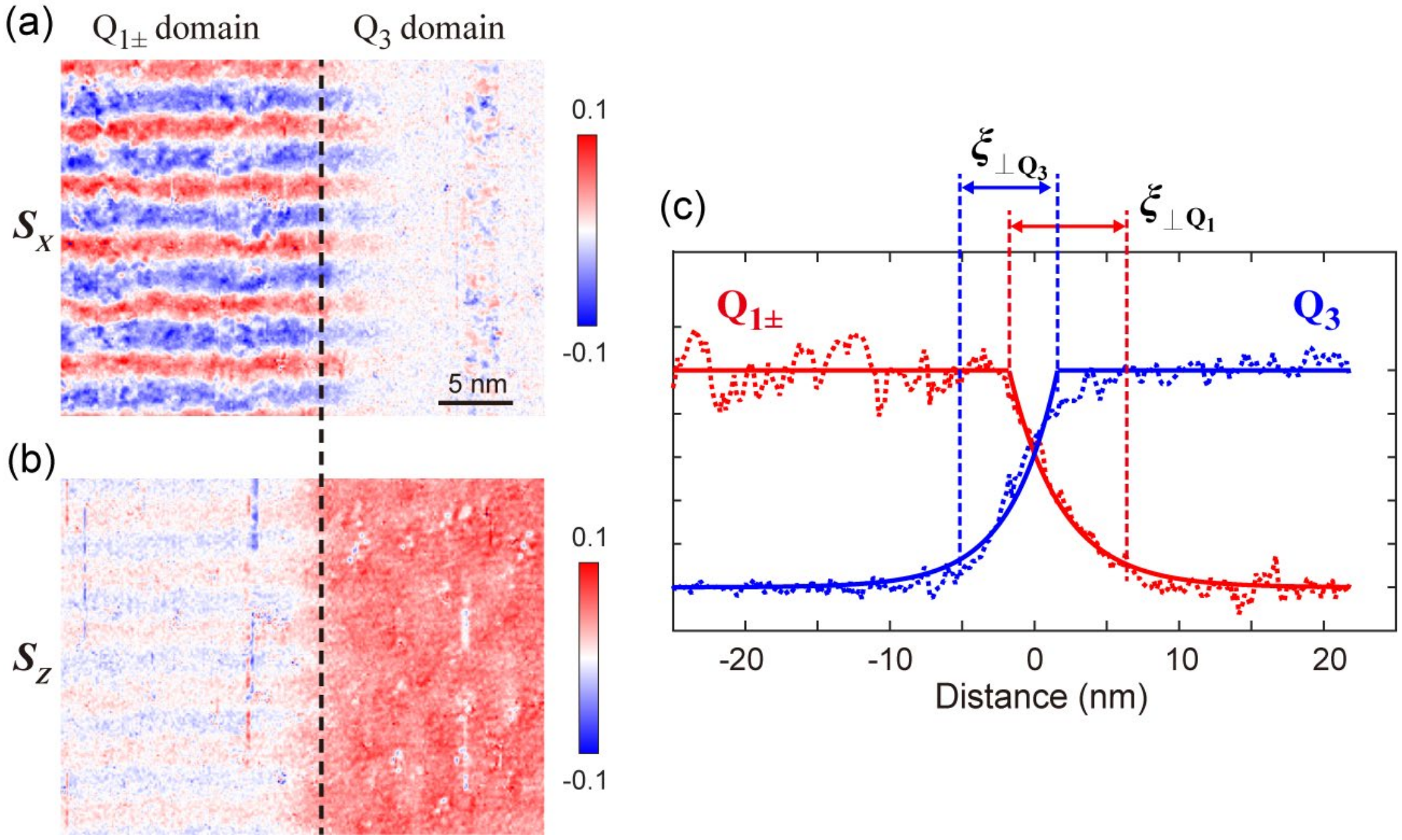

**Fig. S8.** (a,b) $S_X$ and $S_Z$ maps taken around $\boldsymbol{Q}_{1\pm}/\boldsymbol{Q}_3$ domain wall ($I$ = 50pA, $V_b$ = -100 mV). (c) Amplitude profile of the $\boldsymbol{Q}_{1\pm}$ (*IC*-SDW) and $\boldsymbol{Q}_3$ (*C*-SDW) across the domain boundary.

## VIII. Numerical simulation of the electronic states in the anti-phase SDW domain wall.

From the Fermi surface shown Fig. 1(b) and Fig. S9, we observe that for $k_z = 0$, there are three Fermi pockets, with the Γ-pocket electron-like, and the $X$- and $Y$-pockets hole-like. We will now concentrate on the two-dimensional momentum space at kz = 0. This is sufficient for our purpose, although a three-dimensional calculation is also doable. Let the field operator $\psi_k$ describe the low energy states. Since the three pockets are well separated, we can partition the Brillouin zone into patches surrounding the three pockets, and we relabel the states near the three pockets as $\psi_{\Gamma,X,Y}$. We can shift the center of mass of the pockets so that all of them are around the zone center. The SDW order parameter scatters electrons from Γ to $X$ $(Y)$, and vice versa. In this setting, the effective Hamiltonian for a uniform SDW system can be written as, in the momentum space,

$$H = \sum_{k,\alpha=\Gamma,X,Y} \psi^{+}_{\alpha,k} \epsilon^{\alpha}_{\mathrm{k}} \psi_{\alpha,k} - \sum_{k,\alpha=X,Y} \left( \psi^{+}_{\Gamma,\mathrm{k}} m_{\alpha} \sigma_{\alpha} \psi_{\alpha,k} + h.c. \right).$$

Here the spin indices are left implicit, $\epsilon^{\alpha}_{k}$ is the energy dispersion for the $\alpha$-pocket, $\sigma_{\alpha}$ is the Pauli matrix in the spin basis, and $m_{\alpha}$ is the spin scattering potential from the SDW order parameter. This Hamiltonian is very similar to that for a superconductor. To make this point clearer, we rewrite it in the matrix form, in the basis $\psi = (\psi_{\Gamma\uparrow}, \psi_{\Gamma\downarrow}, \psi_{X\uparrow}, \psi_{X\downarrow}, \psi_{Y\uparrow}, \psi_{Y\downarrow})$,

$$H = \sum_{k} \psi^{+}_{k} \begin{pmatrix} \epsilon^{\Gamma}_{\mathrm{k}} & 0 & 0 & m_X & 0 & 0 \\ 0 & \epsilon^{\Gamma}_{\mathrm{k}} & m_X & 0 & 0 & 0 \\ 0 & m_X & \epsilon^{X}_{k} & 0 & 0 & 0 \\ m_X & 0 & 0 & \epsilon^{X}_{k} & 0 & 0 \\ 0 & 0 & 0 & 0 & \epsilon^{Y}_{k} & 0 \\ 0 & 0 & 0 & 0 & 0 & \epsilon^{Y}_{k} \end{pmatrix} \psi_{k},$$

where we assumed the SDW order connects the Γ-X pockets, while the Y-pocket is idle. (The other case of SDW connecting $\Gamma - Y$ pockets can be handled similarly.) We see that the leading 4×4 block of the Hamiltonian matrix is sub-block diagonal, so we concentrate on the sub-block composed of the second and third rows and columns,

$$\begin{pmatrix} \epsilon^{\Gamma}_{k} & m_X \\ m_X & \epsilon^{X}_{k} \end{pmatrix}.$$

The 11-entry of this subblock is electron-like, and the 22-entry is hole-like, mimicking the particle- and hole-component, respectively, of the Bogoliubov-de Gennes (BdG) Hamiltonian in the Nambu space, if the two pockets are exactly nested, namely, $\epsilon^{X}_{k} = -\epsilon^{\Gamma}_{k}$. (If the nesting is not perfect, the center of mass of the BdG band will deviate from zero energy.) The off-diagonal term mimics the local pairing term. Note that here we pair up electron and hole, rather than pairing up electrons. The analogy to superconducting pairing is mathematically valid, and was already pointed out in Rev. Mod. Phys. 60, 209.

To model the case of domain wall, we need to write the Hamiltonian in the real space, which we assume to be given by, on a simple square lattice,

$$H = -\sum_{\alpha,\langle ij \rangle} \left( \psi^{+}_{\alpha,i} t_{\alpha} \psi_{\alpha,j} + h.c. \right) - \sum_{i,\alpha} \psi^{+}_{\alpha,i} (\mu_{\alpha} + m_{\alpha} \sigma_{\alpha}) \psi_{\alpha,i} \,.$$

Here $t_{\alpha} = \pm t$ is the hopping integral on nearest-neighbor bonds for electron/hole pockets, and $\mu_{\alpha} > 0$ ($\mu_{\alpha} < 0$) for electron/hole pocket. In the uniform state, $m_{\alpha}$ is a constant, while in the anti-phase domain wall configuration, we assume it changes as $m_{\alpha}(x) = m_0 \tanh\left(\frac{x}{\xi}\right)$, where $m_0$ is the global amplitude, $x$ is the displacement normal to the domain wall, and $\xi$ is the coherence length, which we assume, without loss of generality, $\xi = 2$ (in units of lattice constant). This Hamiltonian is then similar to that in a superconductor with a Josephson junction, with a phase difference of $\pi$ across the junction. This situation is known to give rise to Andreev bound states, which is an immediate insight from the analogy to superconducting systems. To proceed, we use the above Hamiltonian to calculate the electronic states in the case of anti-phase domain wall. The translation symmetry along the domain wall is preserved, so that the momentum, q, in that direction is conserved. The Hamiltonian can then be partially Fourier-transformed into:

$$H = -\sum_{\alpha,q,i}\left(\psi^{+}_{\alpha,q,i} t_\alpha \psi_{\alpha,q,i+1} + h.c.\right) - \sum_{\alpha,q,i} \psi^{+}_{\alpha,q,i}(\mu_\alpha + 2t_\alpha \cos q + m_\alpha \sigma_\alpha)\psi_{\alpha,q,i}.$$

Here $i$ labels the sites along the x-axis (normal to the domain wall). This Hamiltonian can then be easily diagonalized. The dispersion as a function of q is shown in Fig. 6(a) of the main text, where we set $t = 1$ as the unit of energy, and we set $m_0 = 0.25$ as the strength of the spin scattering potential. On the other hand, we set $\mu_\Gamma = -\mu_X = -\mu_Y = -3$ for the perfect-nesting case, and we set $\mu_\Gamma = -3$ and $\mu_X = \mu_Y = 2.85$ for the case of a smaller electron pocket. Our calculation demonstrates explicitly the existence of a bound state in the case of the anti-phase SDW domain wall.

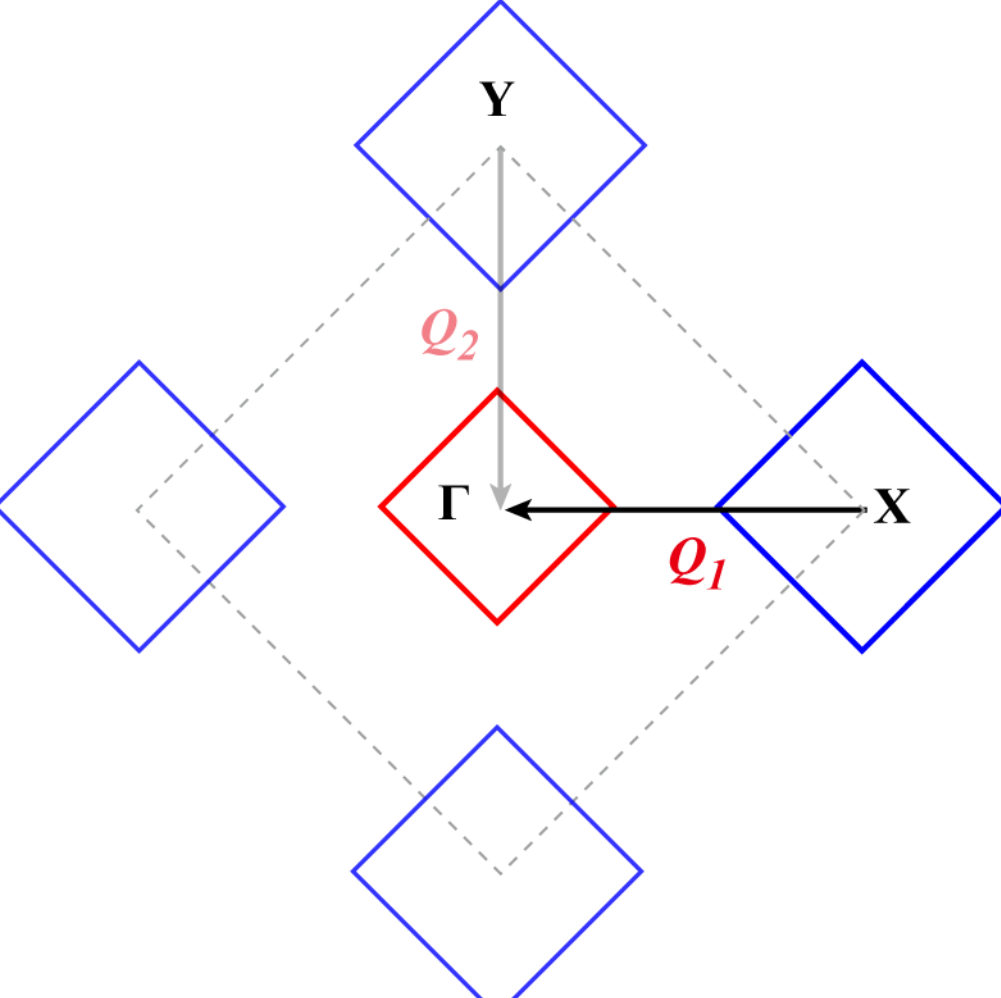

**Fig. S9.** The projected Brillouin zone of Cr. The hole pockets at **X(Y)** points are slightly larger than the electron pocket at **Γ.**